\documentclass[aps,preprint,pra,showpacs,amsfonts,amssymb,amsmath,nofootinbib]{revtex4-1}

\usepackage{subfig}
\usepackage{dsfont}
\usepackage{appendix}
\usepackage{color}
\usepackage{graphicx}
\usepackage{bbold}

\begin{document}

\title{\bf Integral method for the calculation of Hawking radiation in dispersive media\\
I. Symmetric asymptotics}

\author{Scott Robertson}
\affiliation{Laboratoire de Physique Th\'{e}orique, CNRS UMR 8627,\\
B\^{a}timent 210, Universit\'{e} Paris-Sud 11, 91405 Orsay Cedex, France}
\email{scott.robertson@th.u-psud.fr}

\author{Ulf Leonhardt}
\affiliation{Department of the Physics of Complex Systems,\\
Weizmann Institute of Science, Rehovot 76100, Israel}
\email{ulf.leonhardt@weizmann.ac.il}

\begin{abstract}
Hawking radiation has become experimentally testable thanks to the many analogue systems which mimic the effects of the event horizon on wave propagation.  These systems are typically dominated by dispersion, and give rise to a numerically soluble and stable ODE only if the rest-frame dispersion relation $\Omega^{2}(k)$ is a polynomial of relatively low degree.  Here we present a new method for the calculation of wave scattering in a one-dimensional medium of arbitrary dispersion.  It views the wave equation as an integral equation in Fourier space, which can be solved using standard and efficient numerical techniques.
\end{abstract}


\maketitle

\newpage


\section{Introduction
\label{sec:Introduction}}

Hawking radiation -- spontaneous emission from a system in its ground state -- has attracted much interest in recent years.  Ultimately inspired by Hawking's prediction of thermal radiation from a black hole \cite{Hawking-1974,Hawking-1975}, the generality of the radiation was first indicated by Unruh's analogy between black hole spacetime and the effective spacetime of a moving fluid whose flow velocity crosses the speed of sound \cite{Unruh-1981}.  Since then the analogy has been extended to an ever-increasing variety of physical systems \cite{ArtificialBlackHoles,Schutzhold-Unruh,LivingReview}, including water waves \cite{Schutzhold-Unruh-2002,Rousseaux-et-al-2008,Rousseaux-et-al-2010,Weinfurtner-et-al-2011}, light in nonlinear media \cite{Philbin-et-al-2008,Belgiorno-et-al-2010}, and phononic excitations in atomic BEC \cite{Garay-et-al-2000,Garay-et-al-2001,Barcelo-Liberati-Visser-2001-arXiv,Barcelo-Liberati-Visser-2001} and in quantum fluids of light \cite{Solnyshkov-et-al-2011,Gerace-Carusotto-2012}.  It is hoped that these analogue systems, being more accessible than astrophysical black holes, will allow experimental verification of Hawking's prediction in a controlled laboratory setting.  Understanding wave behaviour in such systems -- in particular, the scattering of waves by spatial inhomogeneities in an asymptotically uniform background -- is thus of importance for predicting and interpreting experimental observations.

One major difference between the original black-hole spacetime and the background provided by analogue systems is that the latter are typically dominated by dispersion \cite{Jacobson-1991}, which regularizes the phase singularities at horizons but also complicates the wave equations and makes them less amenable to analytical and numerical techniques.  Apart from time-consuming FDTD wavepacket simulations \cite{Unruh-1995}, existing techniques are restricted to situations in which the background is slowly-varying compared with the scale at which dispersion becomes important \cite{Corley-1998,Leonhardt-Robertson-2012}, in which case analytical methods are applicable; or to dispersion relations which are polynomials of low degree \cite{Corley-Jacobson-1996,Macher-Parentani-2009}, allowing the numerical solution of the ODE in position space provided any exponentially divergent solutions do not severely affect the accuracy of the result\footnote{When the steepness of the change in the background becomes so large that it can be approximated by a step discontinuous function, the solution can be found analytically by matching the plane wave solutions on either side \cite{Corley-1997,Recati-et-al-2009,Finazzi-Parentani-2012}.}.  Given that real dispersion relations are often more complicated than this, and that steepening the variation of the background is the surest way of increasing the radiation rate, we would like to have an efficient numerical algorithm which is not so restricted in its scope.

In this paper, we describe a method of analyzing the scattering properties of waves in dispersive media by working entirely in Fourier space, where the wave equation takes the form of an integral equation.  We shall restrict our attention to the (not uncommon) case of a stationary background which is asymptotically the same both to the left and to the right of the scattering region.  (The generalization to backgrounds with different asymptotic values is treated in a sequel paper, referred to here as Part II.)  The required inputs are the dispersion relation and the half-Fourier transforms of the background.  In Fourier space, the dispersion relation is represented by a multiplicative (rather than differential) operator, and can thus take an arbitrary form.  Steepening of the background corresponds to broadening of its Fourier transform.  Therefore, so long as it is ensured that the integration range and sampling rate are large enough, the method can accommodate a large variety of possible setups.  We emphasise that, although analogue Hawking radiation provides our immediate motivation, the method described in this paper is a general one, applicable to any one-dimensional scattering problem.

The paper is organised as follows.  In Section \ref{sec:Scattering}, we give a brief overview of the connection between scattering amplitudes and Hawking spectra, providing our motivation for the development of the new method.  Section \ref{sec:Wave_equation} describes the wave equation -- in particular the acoustic wave equation considered by Unruh \cite{Unruh-1981,Unruh-1995} -- and shows how it becomes an integral equation under Fourier transformation.  In Section \ref{sec:Integral_equation}, the mathematical theory of the integral method is presented, and an example of its application follows in Section \ref{sec:Application}.  The paper concludes with Section \ref{sec:Conclusion}.


\section{Scattering and Hawking radiation
\label{sec:Scattering}}

To motivate our treatment of scattering, we give here a brief overview of the connection between scattering and Hawking radiation.

\subsection{Basis modes}

Assuming a stationary background which becomes homogeneous asymptotically -- and rejecting on physical grounds all divergent solutions -- a particular solution of a scattering problem can be represented by the amplitudes of the propagating waves (those whose frequencies and wavevectors are {\it real}) far from the scatterer.  In one spatial dimension, there are two natural ways of categorising these far-field propagating waves:
\begin{itemize}
\item {\sc Left and Right}: waves classified according to their position with respect to the scatterer; this distinction is mathematically convenient, as each asymptotic plane wave contributes a singular term -- whose form is determined by which side of the space it lies on -- to the Fourier transform of the overall solution; and
\item {\sc Ingoing and Outgoing}: waves classified according to the direction of their group velocity with respect to the position of the scatterer; this distinction is physically fundamental, as it is equivalent to separating solutions in the asymptotic past from those in the asymptotic future, localising the waves in time rather than in space.
\end{itemize}

In the limit of geometrical optics, any ingoing or outgoing wave must continuously evolve in time to one of the opposite kind\footnote{This is assuming no phase singularities, which occur at a horizon in the absence of dispersion; in particular, it occurs at the event horizon of an astrophysical black hole.  This phenomenon is known as the {\it trans-Planckian problem} \cite{Jacobson-1991,Brout-et-al-Primer}, and is thought to be regularised by as yet unknown dispersive effects at the Planck scale.}.  Thus, the asymptotic plane waves form in/out pairs; and further, for a linear wave equation, the amplitudes of the ingoing and outgoing waves must be linearly related, so that the number of degrees of freedom is simply the number of in/out pairs.  In the case considered in this paper, where the asymptotic values of the background are the same in the left- and right-hand regions, the left- and right-hand sets of wavevector solutions are the same, so that the plane waves can also be thought of as forming left/right pairs with exactly the same frequency and wavevector.

\begin{figure}
\subfloat{\includegraphics[width=0.45\columnwidth]{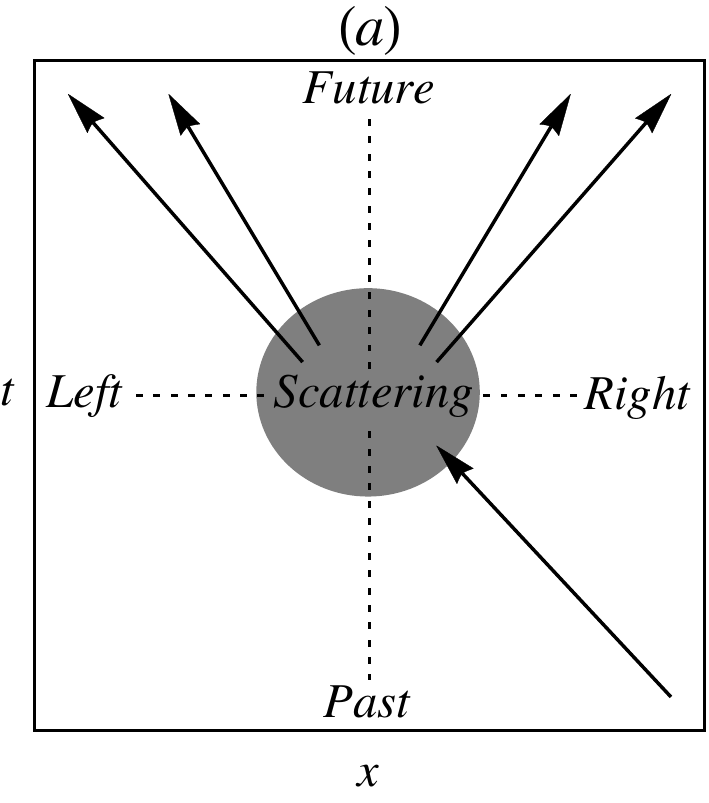}} \, \, \subfloat{\includegraphics[width=0.45\columnwidth]{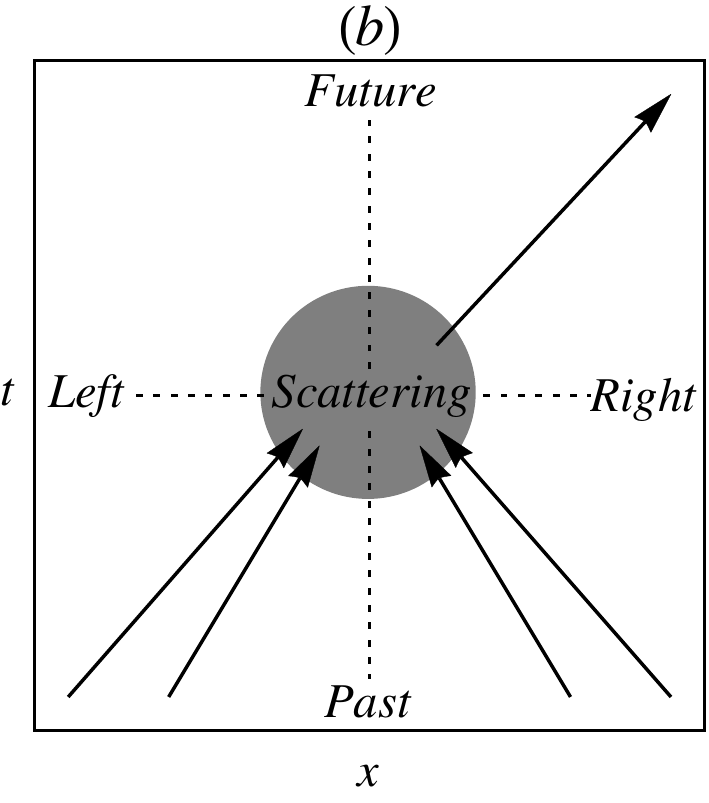}}
\caption{\textsc{In- and Out-Modes} \, Space-time diagrams showing the distinction between $(a)$ an in-mode, with a single plane wave in the asymptotic past; and $(b)$ an out-mode, with a single plane wave in the asymptotic future.  For stationary waves of fixed frequency $\omega$, they can be thought of as the limit of highly peaked wavepackets centred at $\omega$ as the width in Fourier space goes to zero.  The set of all modes of a particular type form a basis for the space of solutions of the scattering problem.
\label{fig:in_and_out_modes}}
\end{figure}

The general solution of a scattering problem can be decomposed into a superposition of ingoing or outgoing modes, which are defined as consisting of a single plane wave in the asymptotic past or future, respectively.  These two sets of modes form bases for the space of solutions, and can be graphically represented using space-time diagrams (see Figure \ref{fig:in_and_out_modes}).  At fixed frequency $\omega$, these modes are stationary, and their ingoing or outgoing nature is best visualised by considering wavepackets formed from a frequency spread highly peaked at $\omega$; the stationary modes themselves are the limiting waveform as the frequency spread is narrowed to a $\delta$ function.

\subsection{Norm conservation and non-uniqueness of the vacuum state}

Scattering solutions are often restricted by conservation laws.  In non-relativistic scattering of particles, for instance, the total number of particles is conserved, while relativistic scattering is subject to such laws as charge conservation.  For waves, there are two conserved quantities:
\begin{itemize}
\item the frequency $\omega$ (whenever the background is stationary); and 
\item a real number called the {\it norm}.
\end{itemize}
Conservation of $\omega$ determines which asymptotic plane waves (labelled by their wavevectors, which form different branches of solutions along which they vary continuously with $\omega$) can scatter into each other, while conservation of norm imposes algebraic relations between the squared magnitudes of the scattering amplitudes \cite{Macher-Parentani-2009}.  If the norms of all plane waves in a given solution have the same sign, its conservation is analogous to non-relativistic scattering and conservation of the number of particles.  If, however, some of the plane waves in a given solution have norms of {\it opposite} sign, its conservation is more akin to charge conservation in relativistic scattering: the total amount of wave energy may well change (in which case a degree of amplification is involved in the scattering process), but the energy difference must be distributed equally between positive and negative norm so that the overall change in norm is zero.

In Quantum Field Theory, waves are interpreted as operator fields, and those of positive and negative norm are multiplied, respectively, by annihilation and creation operators.  Decomposition of out-modes into in-modes can be rearranged to describe decomposition of in-operators into out-operators (see, {\it e.g.}, \cite{Robertson-2012}).  If scattering occurs in which waves are partially converted into waves with opposite norm, then annihilation operators for in-modes will contain creation operators for out-modes, and vice versa.  Given that the quantum vacuum state $\left|0\right>$ is defined mathematically as the zero-eigenvalue eigenstate for all annihilation operators:
\begin{equation}
\hat{a}_{i} \left| 0 \right> = 0 \qquad \forall \, i \,,
\label{eq:vacuum_definition}
\end{equation}
this would mean that the vacuum state for ingoing waves is {\it not} the same as the vacuum state for outgoing waves.  This is the essence of Hawking radiation: in the absence of ingoing particles, outgoing particles are present.  Given that a system is in the in-vacuum, the radiation spectrum of a particular outgoing mode is simply the sum of the squared amplitudes of the opposite-norm ingoing waves occurring in that out-mode.  We can write
\begin{equation}
\phi^{\mathrm{out},+}_{\omega,k_{i}} = \sum_{j} \alpha\left(\omega;k_{i},k_{j}\right) \phi^{\mathrm{in},+}_{\omega,k_{j}} + \sum_{j} \beta\left(\omega;k_{i},k_{j}\right) \phi^{\mathrm{in},-}_{\omega,k_{j}}\,,
\label{eq:out-in_decomp}
\end{equation}
where the modes $\phi_{\omega,k}$ are defined so that their norms are $\pm 1$, the signs being indicated by the superscripts.  The radiation rate per unit frequency of waves on the $k_{i}$-branch is then given by \cite{Corley-Jacobson-1996}
\begin{equation}
\frac{\partial^{2}N_{k_{i}}}{\partial\omega\,\partial t} = \frac{1}{2\pi} \sum_{j} \left| \beta\left(\omega;k_{i},k_{j}\right)\right|^{2}\,.
\label{eq:Hawking_spectrum}
\end{equation}
Furthermore, norm conservation requires \cite{Macher-Parentani-2009,Robertson-2012}
\begin{equation}
\sum_{j} \left| \alpha\left(\omega; k_{i},k_{j}\right) \right|^{2} - \sum_{j} \left| \beta\left( \omega; k_{i},k_{j} \right) \right|^{2} = 1 \,.
\label{eq:norm_conservation}
\end{equation}

Equations (\ref{eq:out-in_decomp}) and (\ref{eq:Hawking_spectrum}) show that calculating the spectrum of Hawking radiation is equivalent to calculating scattering amplitudes for a particular type of scattering process which mixes waves of opposite norm.  From now on, we will focus our attention on finding the scattering matrix $\mathcal{S}$ which relates amplitudes of ingoing waves to those of outgoing waves:
\begin{equation}
\vec{\mathcal{A}}^{\mathrm{out}} = \mathcal{S} \vec{\mathcal{A}}^{\mathrm{in}}\,.
\label{eq:scattering_matrix_definition}
\end{equation}
If the waves to which the amplitudes $\mathcal{A}^{\mathrm{in/out}}$ refer are not themselves normalised, we can define diagonal matrices of normalising prefactors $\hat{\mathcal{N}}^{\mathrm{in/out}}$ such that $\mathcal{A}^{\mathrm{in/out}} = \hat{\mathcal{N}}^{\mathrm{in/out}} \, \mathcal{A}_{N}^{\mathrm{in/out}}$, where $\mathcal{A}_{N}^{\mathrm{in/out}}$ are the amplitudes of the normalised waves.  These are related via the normalised scattering matrix $\mathcal{S}_{N}$:
\begin{alignat}{3}
\vec{\mathcal{A}}^{\mathrm{out}}_{N} = \mathcal{S}_{N} \vec{\mathcal{A}}^{\mathrm{in}}_{N} & \qquad \mathrm{where} & \qquad \mathcal{S}_{N} = \hat{\mathcal{N}}^{\mathrm{out}\,-1} \, \mathcal{S} \, \hat{\mathcal{N}}^{\mathrm{in}} \,.
\label{eq:normalized_scattering_matrix_definition}
\end{alignat}
The elements of $\mathcal{S}_{N}^{-1}$ are the $\alpha$ and $\beta$ coefficients\footnote{In fact, $\mathcal{S}_{N}$ is a member of the indefinite unitary group $U(N_{+},N_{-})$, where $N_{+}$ and $N_{-}$ are, respectively, the number of ingoing (or outgoing) wavevector solutions with positive and negative norm.  If $\eta$ is the diagonal matrix with $N_{+}$ $1$s and $N_{-}$ $-1$s on the diagonal, then $\mathcal{S}_{N}$ obeys $\left[\mathcal{S}_{N}^{T}\right]^{\star} \, \eta \,\mathcal{S}_{N} = \eta$, and hence $\mathcal{S}_{N}^{-1} = \eta \, \left[ \mathcal{S}_{N}^{T} \right]^{\star} \, \eta$.  Thus the coefficients expressing in-modes as sums of out-modes are very closely related to those expressing out-modes as sums of in-modes.} of Eq. (\ref{eq:out-in_decomp}), while those of $\mathcal{S}_{N}$ are the corresponding coefficients with the labels `in' and `out' of Eq. (\ref{eq:out-in_decomp}) switched.  $\mathcal{S}_{N}$ thus gives the scattering amplitudes for each of the in- and out-modes, determines the Hawking spectrum through Eq. (\ref{eq:Hawking_spectrum}) and obeys the norm conservation law (\ref{eq:norm_conservation}).  It is useful to calculate the full matrix $\mathcal{S}_{N}$ even if we are only concerned with certain elements of it (in our case, those relating modes of opposite norm), as it will allow us to check that norm conservation is indeed respected and thus to test the accuracy of the numerical algorithm.


\section{Wave equation
\label{sec:Wave_equation}}

The method to be presented in Section \ref{sec:Integral_equation} solves an integral equation in Fourier space, which is derived by Fourier transforming the wave equation in position space.  To illustrate the transformation, let us perform it in the context of Unruh's original fluid model \cite{Unruh-1981}.  Since this model is concerned with the analogy with waves propagating in spacetime, consider the effective spacetime metric \cite{Unruh-1981,Robertson-2012}
\begin{equation}
ds^{2} = c^{2} dt^{2} - \left( dx - u(x) dt\right)^{2} \,,
\label{eq:spacetime_metric}
\end{equation}
where $c$ is assumed constant and $u(x)$ is assumed independent of time.  This describes motion in a moving fluid, where $c$ is the speed of massless (sound) waves, and $u(x)$ is the local flow velocity of the fluid.  This results, for massless scalar waves, in the wave equation
\begin{equation}
\left( \partial_{t} + \partial_{x}u(x) \right) \left( \partial_{t} + u(x)\partial_{x} \right) \phi - c^{2}\partial_{x}^{2} \phi = 0 \,.
\label{eq:scalar_wave_eqn}
\end{equation}
Introducing dispersion destroys the analogy with a spacetime metric, but it can be achieved by making the speed of sound a function of the wavevector $k$ \cite{Unruh-1995}; or, in position space, $c^{2}$ becomes a differential operator, and if we also impose the harmonic time dependence $\phi(t,x)=e^{-i\omega t} \phi_{\omega}(x)$, the wave equation becomes:
\begin{equation}
\left( -i\omega + \partial_{x}u(x) \right) \left( -i\omega + u(x)\partial_{x} \right) \phi_{\omega} - c^{2}(-i\partial_{x})\partial_{x}^{2} \phi_{\omega} \, = \, 0 \,.
\label{eq:dispersive_wave_eqn}
\end{equation}
Thus we see that, if $c^{2}(k)$ is a polynomial of finite degree, the dispersive wave equation (\ref{eq:dispersive_wave_eqn}) becomes (for each value of $\omega$) an ordinary differential equation, and can in principle be solved numerically.  This cannot be done if $c^{2}(k)$ is more complicated than a polynomial, or if the polynomial is of such a high degree that the solution becomes unstable due to the appearance of exponentially divergent solutions.

Let us now consider the Fourier transformed solution
\begin{equation}
\psi_{\omega}(k) = \int_{-\infty}^{+\infty} \mathrm{d}x \, e^{-ikx} \, \phi_{\omega}(x) \,.
\label{eq:FT}
\end{equation}
In the wave equation (\ref{eq:dispersive_wave_eqn}), spatial derivatives $-i\partial_{x}$ are replaced by the wavevector $k$; in particular, the differential operator $c^{2}(-i\partial_{x})$ is replaced by the multiplicative operator $c^{2}(k)$. In turn, multiplicative operators in position space -- here these are due to the flow velocity $u(x)$ -- give rise to convolutions in Fourier space, since the Fourier transform of a product of functions is the convolution of their Fourier transforms\footnote{It is also possible to replace $x$ by $i\partial_{k}$, treating $u(x)$ as a differential operator in Fourier space which can be truncated at a low degree to find approximate solutions; this is the approach taken when the background is assumed to be slowly-varying in relation to the length scale at which dispersion becomes important, as in \cite{Corley-1998,Leonhardt-Robertson-2012}.}.  The occurrence of further spatial derivatives in the terms containing $u(x)$, plus the multiplication of two factors of $u(x)$, means that the final result is more complicated than a straightforward convolution, though it is still an integral that contains the Fourier transforms of the flow velocity, $\mathcal{F}[u](k)$, and of its square, $\mathcal{F}[u^{2}](k)$.  The complete Fourier transform of Eq. (\ref{eq:dispersive_wave_eqn}) is
\begin{equation}
g_{\omega}(k) \psi_{\omega}(k) + \int_{-\infty}^{+\infty} K_{\omega}(k,k^{\prime}) \psi_{\omega}(k^{\prime}) \, \mathrm{d}k^{\prime} = 0\,,
\label{eq:FT_integral_eqn}
\end{equation}
where we have defined
\begin{eqnarray}
g_{\omega}(k) & = & c^{2}(k)k^{2} - \omega^{2} \,,
\label{eq:fluid_dispersion}\\
K_{\omega}(k,k^{\prime}) & = & \frac{1}{2\pi} \left[ 2 \omega k\, \mathcal{F}[u](k-k^{\prime}) + i\omega\, \mathcal{F}[\partial_{x}u](k-k^{\prime}) \right. \nonumber \\
& & \qquad \qquad \left. - k^{2}\, \mathcal{F}[u^{2}](k-k^{\prime}) - ik\, \mathcal{F}[\partial_{x}u^{2}](k-k^{\prime}) \right] \nonumber \\
& = & \frac{1}{2\pi} \left[ \omega \left(k+k^{\prime}\right) \mathcal{F}[u](k-k^{\prime}) - k k^{\prime} \mathcal{F}[u^{2}](k-k^{\prime}) \right] \, .
\label{eq:fluid_kernel}
\end{eqnarray}
Equation (\ref{eq:dispersive_wave_eqn}) was used here only as an illustration.  An analogous procedure can be performed for any linear wave equation: the position-independent part becomes a straightforward multiplicative term, and all the position-dependent terms are absorbed into the kernel of the integral operator, functions of position being replaced by convolution-like integrals.


\section{Solving the integral equation
\label{sec:Integral_equation}}

\subsection{The problem}

As discussed in Section \ref{sec:Wave_equation}, Fourier transforming the wave equation results in an equation whose general form is
\begin{equation}
g(k) \psi(k) + \int_{-\infty}^{+\infty} K(k,k^{\prime}) \psi(k^{\prime}) \mathrm{d}k^{\prime} = 0\,,
\label{eq:general_integral_equation}
\end{equation}
where for clarity we have suppressed the explicit $\omega$-dependence of Eq. (\ref{eq:FT_integral_eqn}).  This is a homogeneous integral equation of the third kind \cite{Bart-Warnock-1973}.  Homogeneity is a necessary property, since we must have the trivial solution $\psi(k) = 0$.  The equation is said to be {\it of the third kind} because $g(k)$ vanishes at certain real values of $k$, rendering the operator acting on $\psi(k)$ singular.  To see this, consider Eq. (\ref{eq:general_integral_equation}) as the continuous limit of the discretised equation
\begin{equation}
\sum_{m}\left[ g\left(k_{n}\right) \delta_{nm} + \Delta k \cdot K\left(k_{n},k_{m}\right)\right] \psi\left(k_{m}\right) = 0\,,
\label{eq:discretized_integral_equation}
\end{equation}
which is how one would model Eq. (\ref{eq:general_integral_equation}) numerically.  Notice that only $K\left(k_{n},k_{m}\right)$ is multiplied by $\Delta k$, so that, as $\Delta k$ is made smaller, the elements of the matrix multiplying $\psi\left(k_{m}\right)$ are dominated by the values of $g\left(k_{n}\right)$ on the diagonal.  Without being mathematically rigorous, the determinant of this matrix is essentially the product of the values of $g\left(k_{n}\right)$.  Thus, if $g(k)=0$ anywhere in the integration interval, the operator becomes singular.  Indeed, this is a necessary condition if we are to have non-trivial, non-unique solutions of a linear homogeneous equation.

Since we are considering equal asymptotic values of the background, it can be described by some constant value plus a spatially dependent term which vanishes asymptotically.  The constant value gives rise to a $\delta$ function in the Fourier transform of the background, such that the kernel $K(k,k^{\prime})$ contains a term proportional to $\delta(k-k^{\prime})$.  This is easily integrated and incorporated into the first term of Eq. (\ref{eq:general_integral_equation}), leaving the equation in exactly the same form.  (For the example of Eqs. (\ref{eq:fluid_dispersion}) and (\ref{eq:fluid_kernel}), this amounts to redefining $g_{\omega}(k) \rightarrow c^{2}(k)k^{2}-\left(\omega-u_{0}k\right)^{2}$, $u(x) \rightarrow u(x)-u_{0}$ and $u^{2}(x) \rightarrow u^{2}(x)-u_{0}^{2}$.)  Therefore, we can assume that $K(k,k^{\prime})$ is a smooth function of $k$ and $k^{\prime}$.

Equation (\ref{eq:general_integral_equation}) is exactly the type of integral equation studied by Bart and Warnock \cite{Bart-Warnock-1973}.  First we note that, setting $K(k,k^{\prime})=0$ so that the background is constant, the resulting equation $g(k)\psi(k)=0$ states simply that $\psi(k)$ can only be non-zero at the roots of $k$.  Since the roots are typically a collection of points of measure zero, the only non-trivial contribution to the Fourier transform is a collection of independent $\delta$ functions at these points.  The roots of $g(k)$, then, are simply the solutions of the dispersion relation in the asymptotic regions.  We use $N$ to denote the number of real roots -- which correspond to propagating waves -- and label them with the subscripts $i$ and $j$ (as opposed to the subscripts $n$ and $m$, which we use as in Eq. (\ref{eq:discretized_integral_equation}) for the discretised values of $k$ in the integration region).  In general, $\psi(k)$ consists of $\delta$ functions and poles at these points.  Bart and Warnock showed how to find a linear relationship between the coefficients of the $\delta$ functions and those of the poles by rearranging Eq. (\ref{eq:general_integral_equation}) such that the operator becomes non-singular and thus invertible.  Inspired by Bart and Warnock, we follow their analysis closely, but it is modified so as to aim at a relationship between the asymptotic plane waves rather than between $\delta$ functions and poles.  This also has the advantage over Bart and Warnock's method of being generalisable to the case of unequal asymptotic values of the background.  (Such a generalisation is the subject of Part II.)


\subsection{Half-Fourier transforms}

To solve Eq. (\ref{eq:general_integral_equation}), we begin (differently from Bart and Warnock \cite{Bart-Warnock-1973}) by splitting $\psi(k)$ into two half-Fourier transforms of the left- and right-hand sides:
\begin{alignat}{2}
\psi^{L}(k) = \int_{-\infty}^{0} \phi(x)e^{-ikx} \mathrm{d}x\,, & \qquad \psi^{R}(k) = \int_{0}^{+\infty} \phi(x)e^{-ikx} \mathrm{d}x\,.
\label{eq:half-FTs}
\end{alignat}
Since $\phi(x)$ is assumed to be asymptotically bounded, $\psi^{L}(k)$ must be analytic and go to zero at least as fast as $k^{-1}$ in the upper-half complex $k$-plane, while $\psi^{R}(k)$ behaves similarly in the lower-half $k$-plane.  Assuming sufficient smoothness of the background at $x=0$, the kernel $K(k,k^{\prime})$ can also be split in such a manner with respect to the variable $k^{\prime}$ (see Appendix \ref{app:splitting_kernel}); {\it i.e.} we can write
\begin{equation}
K(k,k^{\prime}) = K_{L}(k,k^{\prime}) + K_{R}(k,k^{\prime})
\label{eq:splitting_kernel}
\end{equation}
where $K_{L}(k,k^{\prime})$ is analytic and approaches zero at least as fast as $1/k^{\prime}$ in the {\it lower} half $k^{\prime}$-plane, and $K_{R}(k,k^{\prime})$ behaves similarly in the {\it upper} half $k^{\prime}$-plane; we attach $L$ and $R$ as subscripts, as opposed to superscripts in Eqs. (\ref{eq:half-FTs}), to indicate this difference.  The advantage of this decomposition is two-fold.  Firstly, the product of two functions possessing these analyticity properties on the {\it same} half plane vanishes upon integration, since the integration contour can be closed on the half plane where analyticity holds.  Using this fact upon substitution of the half-Fourier transforms into Eq. (\ref{eq:general_integral_equation}), it becomes
\begin{equation}
g(k) \big( \psi^{L}(k) + \psi^{R}(k) \big) + \int_{-\infty}^{+\infty} K_{L}(k,k^{\prime}) \psi^{L}(k^{\prime}) \mathrm{d}k^{\prime} + \int_{-\infty}^{+\infty} K_{R}(k,k^{\prime}) \psi^{R}(k^{\prime}) \mathrm{d}k^{\prime} = 0\,.
\label{eq:left-right_integral_equation}
\end{equation}
The second advantage of decompositions (\ref{eq:half-FTs}) and (\ref{eq:splitting_kernel}) is that we can explicitly evaluate the Hilbert transform of a half-plane analytic function that vanishes asymptotically at least as fast as $1/k$: it is simply half of the value associated with the Cauchy residue theorem.  This is useful when the half-Fourier transforms of the propagating waves are introduced, since they contain poles on the real $k$-axis and generate precisely such Hilbert transforms.


\subsection{Extraction of asymptotic plane waves}

To solve Eq. (\ref{eq:left-right_integral_equation}), we decompose $\psi^{L}(k)$ and $\psi^{R}(k)$ into a singular part and a regular part.  The singular part derives from the behaviour at infinity, which is due to the half-Fourier transforms of the asymptotic plane waves.  For a plane wave $\mathrm{exp}\left(ik_{w}x\right)$, these are
\begin{subequations}
\begin{eqnarray}
\frac{1}{2\pi} \int_{-\infty}^{0} e^{ik_{w}x} e^{-ikx} \mathrm{d}x & = & \frac{1}{2} \, \delta\left( k-k_{w} \right) - \frac{1}{2\pi\,i} \, \mathcal{P} \frac{1}{k-k_{w}} \,, \\
\frac{1}{2\pi} \int_{0}^{+\infty} e^{ik_{w}x} e^{-ikx} \mathrm{d}x & = & \frac{1}{2} \, \delta\left( k-k_{w} \right) + \frac{1}{2\pi\,i} \, \mathcal{P} \frac{1}{k-k_{w}} \,,
\end{eqnarray}\label{eqs:FT_plane_wave}\end{subequations}
where $\mathcal{P}$ stands for the principal part, to be taken upon integration.  Accordingly, we write\footnote{The amplitudes $\mathcal{A}^{L/R}_{j}$ introduced here are the coefficients of the unnormalised plane waves.  Normalisation is effected by the prefactors $\left| g^{\prime}\left(k_{j}\right) \right|^{-1/2}$.  These appear on the diagonal of $\hat{\mathcal{N}}^{\mathrm{in/out}}$ in Eq. (\ref{eq:normalized_scattering_matrix_definition}), used to transform the scattering matrix for unnormalised waves $\mathcal{S}$ into that for normalised waves $\mathcal{S}_{N}$.}
\begin{subequations}
\begin{eqnarray}
\psi^{L}(k) & = & \alpha^{L}(k) + \sum_{j=1}^{N_{L}} \mathcal{A}_{j}^{L} \left[ \frac{1}{2} \, \delta\left( k-k_{j}^{(L)} \right) - \frac{1}{2\pi\,i} \, \mathcal{P} \frac{1}{k-k_{j}^{(L)}} \right] \,, \\
\psi^{R}(k) & = & \alpha^{R}(k) + \sum_{j=1}^{N_{R}} \mathcal{A}_{j}^{R} \left[ \frac{1}{2} \, \delta\left( k-k_{j}^{(R)} \right) + \frac{1}{2\pi\,i} \, \mathcal{P} \frac{1}{k-k_{j}^{(R)}} \right] \,,
\end{eqnarray}\label{eqs:FT_psi}\end{subequations}
Here, $\alpha^{L}(k)$ and $\alpha^{R}(k)$ are the remaining regular parts of $\psi^{L}(k)$ and $\psi^{R}(k)$; they encode the wave behaviour in the scattering region.  Substituting Eqs. (\ref{eqs:FT_psi}) into Eq. (\ref{eq:left-right_integral_equation}) and using the $k^{\prime}$-analyticity properties of $K_{L/R}(k,k^{\prime})$ to explicitly evaluate the Hilbert transforms that occur on their multiplication by poles, we obtain an integral equation for $\alpha(k)=\alpha^{L}(k)+\alpha^{R}(k)$:
\begin{multline}
g(k) \alpha(k) + \int_{-\infty}^{+\infty} K(k,k^{\prime}) \alpha(k^{\prime}) \, \mathrm{d}k^{\prime} \\
+ \sum_{j=1}^{N} \mathcal{A}_{j}^{L} \left[ -\frac{1}{2\pi\,i} \, \frac{g(k)}{k-k_{j}} + K_{L}(k,k_{j}) \right] + \sum_{j=1}^{N} \mathcal{A}_{j}^{R} \left[ \frac{1}{2\pi\,i} \, \frac{g(k)}{k-k_{j}} + K_{R}(k,k_{j}) \right] = 0\,.
\label{eq:alpha_integral_equation}
\end{multline}


\subsection{Regularisation of the equation}

Although Eq. (\ref{eq:alpha_integral_equation}) differs from Eq. (\ref{eq:general_integral_equation}) in that it is inhomogeneous, it remains formally singular due to multiplication by $g(k)$ in its first term.  However, unlike $\psi(k)$ in Eq. (\ref{eq:general_integral_equation}), we have defined $\alpha(k)$ to be regular, and therefore, as $k \to k_{j}$ where $g(k_{j})=0$, the first term of Eq. (\ref{eq:alpha_integral_equation}) must vanish.  This leads to a set of regularity conditions for $\alpha(k)$, one for each root of $g(k)$:
\begin{multline}
\int_{-\infty}^{+\infty} K(k_{i},k^{\prime}) \alpha(k^{\prime}) \mathrm{d}k^{\prime} + \sum_{j=1}^{N} \mathcal{A}^{L}_{j} \left[ -\frac{1}{2\pi\,i} \, g^{\prime}(k_{i}) \delta_{ij} + K_{L}(k_{i},k_{j}) \right] \\
+ \sum_{j=1}^{N} \mathcal{A}^{R}_{j} \left[ \frac{1}{2\pi\,i} \, g^{\prime}(k_{i}) \delta_{ij} + K_{R}(k_{i},k_{j}) \right] = 0 \, .
\label{eq:regularity_conditions}
\end{multline}
We are free to add multiples of Eq. (\ref{eq:regularity_conditions}) to the integral equation (\ref{eq:alpha_integral_equation}), and we note that if we subtract one of them, then as $k \to k_{i}$ {\it each} term of the equation vanishes like $k-k_{i}$.  It then becomes possible to divide by $k-k_{i}$, leaving a non-trivial equation.  In order that all the zeros of $g(k)$ may be removed, we follow Bart and Warnock by defining a set of functions $f_{j}(k)$ such that $f_{j}\left(k_{i}\right)=\delta_{ij}$.  Such a set of functions is easily constructed \footnote{Note that we are assuming $g^{\prime}(k_{i}) \neq 0$ for all $i$.  This condition fails at particular frequencies corresponding to the extrema of the dispersion relation $\omega(k)$, where two real roots coalesce and become complex.  These mark the boundaries between regions of the spectrum where the number of real solutions of the dispersion relation differs, and which must be treated separately in numerical calculations.} from $g(k)$:
\begin{equation}
f_{j}(k) = \frac{g(k)}{\left(k-k_{j}\right) \, g^{\prime}\left(k_{j}\right)} \,.
\label{eq:unit_functions}
\end{equation}
From Eq. (\ref{eq:alpha_integral_equation}) we then subtract the sum of the $f_{j}(k)$ multiplied by the corresponding regularity condition (\ref{eq:regularity_conditions}), thus making each term of the equation vanish like $k-k_{i}$ at each of the roots of $g(k)$.  This allows us to divide by $g(k)$, yielding
\begin{equation}
\alpha(k) + \int_{-\infty}^{+\infty} \bar{K}(k,k^{\prime}) \alpha(k^{\prime}) \, \mathrm{d}k^{\prime} + \sum_{j=1}^{N} \mathcal{A}^{L}_{j} \bar{K}_{L}(k,k_{j}) + \sum_{j=1}^{N} \mathcal{A}^{R}_{j} \bar{K}_{R}(k,k_{j}) = 0 \,.
\label{eq:alpha_integral_equation_nonsingular}
\end{equation}
Overbars have been used to indicate subtraction of the regularity conditions (\ref{eq:regularity_conditions}) multiplied by the functions (\ref{eq:unit_functions}), followed by division by $g(k)$.  Because of the definitions we have used for the functions (\ref{eq:unit_functions}), this is equivalent to first dividing by $g(k)$ and then subtracting all the poles on the real axis.  So, for a general function $F(k)$, we have
\begin{equation}
\bar{F}(k) = \frac{F(k)}{g(k)} - \sum_{j=1}^{N} \frac{F(k_{j})}{(k-k_{j}) \, g^{\prime}(k_{j})} \,.
\label{eq:overbar_definition}
\end{equation}


\subsection{Solving for the plane wave amplitudes}

Equation (\ref{eq:alpha_integral_equation_nonsingular}) is now an inhomogeneous integral equation of the {\it second} kind for $\alpha(k)$, so-called because the operator acting on $\alpha(k)$ is invertible.  In its discretized form (analogous to Eq. (\ref{eq:discretized_integral_equation})), it becomes
\begin{equation}
\sum_{m} \left[ \delta_{nm} + \Delta k \cdot \bar{K}(k_{n},k_{m}) \right] \alpha(k_{m}) = -\sum_{j} \bar{K}_{L}(k_{n},k_{j}) \mathcal{A}^{L}_{j} - \sum_{j} \bar{K}_{R}(k_{n},k_{j}) \mathcal{A}^{R}_{j} \,.
\end{equation}
The matrix multiplying $\alpha$ has an inverse, $V=\left[\delta_{nm} + \Delta k \cdot \bar{K}\left(k_{n},k_{m}\right) \right]^{-1}$, so that
\begin{equation}
\alpha(k_{n}) = -\sum_{m}\sum_{j} V_{nm} \bar{K}_{L}(k_{m},k_{j}) \mathcal{A}^{L}_{j} - \sum_{m}\sum_{j} V_{nm} \bar{K}_{R}(k_{m},k_{j}) \mathcal{A}^{R}_{j} \,,
\label{eq:alpha_matrix_equation}
\end{equation}
Similarly, in the continuous limit, there is an inverse kernel $V(k,k^{\prime})$ such that
\begin{equation}
\alpha(k) = -\int_{-\infty}^{+\infty} V(k,k^{\prime}) \left( \sum_{j} \mathcal{A}^{L}_{j} \bar{K}_{L}(k^{\prime},k_{j}) + \mathcal{A}^{R}_{j} \bar{K}_{R}(k^{\prime},k_{j}) \right) \mathrm{d}k^{\prime} \,.
\label{eq:alpha_integral_solution}
\end{equation}

Finally, we note that, while we subtracted the regularity conditions (\ref{eq:regularity_conditions}) from Eq. (\ref{eq:alpha_integral_equation}) to form a non-singular equation, we have yet to enforce those conditions to ensure that the transformed equation is equivalent to the original one.  So while Eqs. (\ref{eq:alpha_matrix_equation}) and (\ref{eq:alpha_integral_solution}) on their own have $2N$ degrees of freedom, the $N$ regularity conditions (\ref{eq:regularity_conditions}) reduce these to $N$ degrees of freedom -- exactly the number of in/out pairs and hence the dimension of the space of (asymptotically bounded) solutions.  Substituting the form of $\alpha$ given by Eq. (\ref{eq:alpha_integral_solution}) into Eq. (\ref{eq:regularity_conditions}), we have
\begin{multline}
\sum_{j} \left[ -\frac{1}{2\pi\,i} \, g^{\prime}(k_{i}) \delta_{ij} + K_{L}(k_{i},k_{j}) - \int_{-\infty}^{+\infty} \mathrm{d}k \int_{-\infty}^{+\infty} \mathrm{d}k^{\prime} K(k_{i},k) V(k,k^{\prime}) \bar{K}_{L}(k^{\prime},k_{j}) \right] \mathcal{A}^{L}_{j} \\
+ \sum_{j} \left[ \frac{1}{2\pi\,i} \, g^{\prime}(k_{i}) \delta_{ij} + K_{R}(k_{i},k_{j}) - \int_{-\infty}^{+\infty} \mathrm{d}k \int_{-\infty}^{+\infty} \mathrm{d}k^{\prime} K(k_{i},k) V(k,k^{\prime}) \bar{K}_{R}(k^{\prime},k_{j}) \right] \mathcal{A}^{R}_{j} \\
= 0 \,,
\end{multline}
or
\begin{equation}
\mathcal{M}_{L} \vec{\mathcal{A}}^{L} + \mathcal{M}_{R} \vec{\mathcal{A}}^{R} = 0
\label{eq:left-right_matrix_equation}
\end{equation}
where
\begin{multline}
\left[ \mathcal{M}_{L} \right]_{ij} \, = \, -\frac{1}{2\pi\,i} g^{\prime}(k_{i}) \delta_{ij} + K_{L}(k_{i},k_{j}) - \int_{-\infty}^{+\infty} \mathrm{d}k \int_{-\infty}^{+\infty} \mathrm{d}k^{\prime} \, K(k_{i},k) V(k,k^{\prime}) \bar{K}_{L}(k^{\prime},k_{j}) \,,\\
\left[ \mathcal{M}_{R} \right]_{ij} \, = \, \frac{1}{2\pi\,i} g^{\prime}(k_{i}) \delta_{ij} + K_{R}(k_{i},k_{j}) - \int_{-\infty}^{+\infty} \mathrm{d}k \int_{-\infty}^{+\infty} \mathrm{d}k^{\prime} \, K(k_{i},k) V(k,k^{\prime}) \bar{K}_{R}(k^{\prime},k_{j}) \,.
\label{eq:left-right_matrices_integral}
\end{multline}
In discretized form, the matrices become
\begin{eqnarray}
\left[ \mathcal{M}_{L} \right]_{ij} & = & -\frac{1}{2\pi\,i} g^{\prime}(k_{i}) \delta_{ij} + K_{L}(k_{i},k_{j}) - \sum_{n,m} \Delta k \cdot K(k_{i},k_{n}) V_{nm} \bar{K}_{L}(k_{m},k_{j}) \,, \nonumber \\
\left[ \mathcal{M}_{R} \right]_{ij} & = & \frac{1}{2\pi\,i} g^{\prime}(k_{i}) \delta_{ij} + K_{R}(k_{i},k_{j}) - \sum_{n,m} \Delta k \cdot K(k_{i},k_{n}) V_{nm} \bar{K}_{R}(k_{m},k_{j}) \,.
\label{eq:left-right_matrices_discretized}
\end{eqnarray}


\subsection{Rearranging into the in- and out-bases}

Equations (\ref{eq:left-right_matrix_equation})-(\ref{eq:left-right_matrices_discretized}) describe the linear relationship between the amplitudes of the asymptotic plane waves, as desired.  Because we have restricted our attention to a system whose background is the same in each of the asymptotic regions, the set of plane waves is the same in each region, so that the relationship can be straightforwardly expressed between the left- and right-hand sides, as it is here.  It can also be rearranged into a relation between ingoing and outgoing modes, as we shall now show.

Each wavevector is ingoing in one asymptotic region and outgoing in the other.  This can be specified using the additional variable $s_{j}$, which we set to $-1$ if $k_{j}$ is outgoing to the left (negative) side and $+1$ if $k_{j}$ is outgoing to the right (positive) side.  Now introduce the projection operators $Q^{-}$ and $Q^{+}$: $Q^{-}$ projects out the wave amplitudes whose corresponding wavevectors have $s=-1$, and similarly $Q^{+}$ for $s=+1$.  In matrix form, they are diagonal matrices whose entries are $1$ or $0$ according to whether the corresponding wavevectors have the appropriate value of $s$.  The amplitudes in the in- and out-bases are then
\begin{alignat}{2}
\vec{\mathcal{A}}^{\mathrm{out}} = Q^{-} \vec{\mathcal{A}}^{L} + Q^{+} \vec{\mathcal{A}}^{R} \,, & \qquad \vec{\mathcal{A}}^{\mathrm{in}} = Q^{+} \vec{\mathcal{A}}^{L} + Q^{-} \vec{\mathcal{A}}^{R} \,.
\label{eq:left-right_to_in-out}
\end{alignat}
Defining $\mathcal{M}=-\left( \mathcal{M}_{R} \right)^{-1} \mathcal{M}_{L}$ so that Eq. (\ref{eq:left-right_matrix_equation}) is equivalent to $\vec{\mathcal{A}}^{R}=\mathcal{M}\vec{\mathcal{A}}^{L}$, and using the following properties of the projection operators:
\begin{subequations}
\begin{eqnarray}
Q^{-}+Q^{+} & = & \mathbb{1} \,, \\
Q^{-} Q^{+} = Q^{+} Q^{-} & = & 0 \,, \\
Q^{-} Q^{-} & = & Q^{-} \,, \\
Q^{+} Q^{+} & = & Q^{+} \,,
\end{eqnarray}
\end{subequations}
we have:
\begin{eqnarray}
\vec{\mathcal{A}}^{\mathrm{out}} & = & Q^{-} \vec{\mathcal{A}}^{L} + Q^{+} \vec{\mathcal{A}}^{R} \nonumber \\
& = & Q^{-} \mathcal{M}^{-1} \left( Q^{-}+Q^{+} \right) \vec{\mathcal{A}}^{R} + Q^{+} \mathcal{M} \left( Q^{-}+Q^{+} \right) \vec{\mathcal{A}}^{L} \nonumber \\
& = & \left( Q^{-}\mathcal{M}^{-1}Q^{-} + Q^{+}\mathcal{M}Q^{+} \right)\left( Q^{-}\vec{\mathcal{A}}^{R} + Q^{+}\vec{\mathcal{A}}^{L} \right) \nonumber \\
& & \qquad + \left( Q^{-}\mathcal{M}^{-1}Q^{+} + Q^{+}\mathcal{M}Q^{-} \right) \left( Q^{+}\vec{\mathcal{A}}^{R} + Q^{-}\vec{\mathcal{A}}^{L} \right) \nonumber \\
& = & \left( Q^{-}\mathcal{M}^{-1}Q^{-} + Q^{+}\mathcal{M}Q^{+} \right) \vec{\mathcal{A}}^{\mathrm{in}} + \left( Q^{-}\mathcal{M}^{-1}Q^{+} + Q^{+}\mathcal{M}Q^{-} \right) \vec{\mathcal{A}}^{\mathrm{out}}
\end{eqnarray}
or
\begin{equation}
\left( \mathbb{1} - Q^{-}\mathcal{M}^{-1}Q^{+} - Q^{+}\mathcal{M}Q^{-} \right) \vec{\mathcal{A}}^{\mathrm{out}} = \left( Q^{-}\mathcal{M}^{-1}Q^{-} + Q^{+}\mathcal{M}Q^{+} \right) \vec{\mathcal{A}}^{\mathrm{in}} \,.
\end{equation}
This is the equivalent relation expressed in terms of the in- and out-bases.  Comparing with Eq. (\ref{eq:scattering_matrix_definition}), we see that the scattering matrix is given by
\begin{equation}
\mathcal{S} = \left[ \mathbb{1} - Q^{-}\mathcal{M}^{-1}Q^{+} - Q^{+}\mathcal{M}Q^{-} \right]^{-1} \left[ Q^{-}\mathcal{M}^{-1}Q^{-} + Q^{+}\mathcal{M}Q^{+} \right] \,.
\label{eq:scattering_matrix_soln}
\end{equation}


\subsection{Numerics}

While Eqs. (\ref{eq:left-right_matrices_integral}) give the exact relationship between the wave amplitudes, they do not easily yield analytic solutions, mainly due to the appearance of the inverse kernel $V(k,k^{\prime})$, which is generally very difficult to compute analytically.  However, upon discretization in Eqs. (\ref{eq:left-right_matrices_discretized}), the inverse kernel becomes an inverse matrix, which can be efficiently calculated using standard numerical algorithms.

Numerical techniques are also useful in calculating the integrals, which become sums in the discretised form (\ref{eq:left-right_matrices_discretized}).  The behaviour of the integrands should determine the parameters of the discretisation: the spacing $\Delta k$, the limits $\pm k_{\mathrm{max}}$ and the number of points on the grid $M=2\,k_{\mathrm{max}}/\Delta k$.)  Looking more closely at the nature of the terms in the integrand, we find that while $V(k,k^{\prime})$ and $\bar{K}_{L/R}(k,k^{\prime})$ go as $(k-k^{\prime})^{-1}$ when $|k-k^{\prime}|$ is large, $K(k,k^{\prime})$ typically decays much faster than this for well-behaved velocity profiles\footnote{This is not so when the asymptotic values of the background are different, as is shown in Part II.}.  So, assuming a smooth background, the form of the integrand is determined mainly by the form of $K(k,k^{\prime})$, which in turn is essentially given by the Fourier transform of the background.  Since both the spacing $\Delta k$ and the limits $\pm k_{\mathrm{max}}$ vary in proportion to the width of the Fourier transform, the number of points $M$ does not vary and the time taken to perform the calculations remains the same over a range of widths.  This breaks down only when the background is well inside the slowly-varying regime and hence has a narrow Fourier transform, for while the spacing $\Delta k$ must continue to decrease as this width decreases, the solutions of the dispersion relation are fixed and must remain comfortably inside the range of integration, so that the limits $\pm k_{\mathrm{max}}$ become constant; the net effect is that $M$ has to increase.  Fortuitously, this is precisely the limit in which analytical techniques become valid \cite{Corley-1998,Leonhardt-Robertson-2012}, so the increase in computation required here is not a great loss.


\section{Application
\label{sec:Application}}

Here we shall justify the method presented in Section \ref{sec:Integral_equation} with results for concrete examples.  We shall continue in the spirit of Unruh's flowing fluid analogy, so that the system is described by Eq. (\ref{eq:dispersive_wave_eqn}) and is determined by the choice of dispersion relation $c(k)$ and velocity profile $u(x)$.  We shall test two dispersion relations: one a low-degree polynomial that can also be solved by the standard ODE in position space, allowing comparison between the two methods; the other a realistic dispersion relation that cannot be properly treated in this way.  We shall also use two velocity profiles differing in their rates of variation, testing the method in the standard Hawking (slowly-varying) regime and when the variation is too rapid for Hawking's prediction to be applicable.


\subsection{Dispersion relation}

Let us consider the familiar example of waves on the surface of water, which, if the effects of surface tension can be neglected, obey the dispersion relation \cite{Landau-Lifshitz-FM}
\begin{equation}
c^{2}(k) = \frac{g}{k} \, \mathrm{tanh} \left( h k \right) \,.
\label{eq:water_dispersion}
\end{equation}
Here, $g$ is the gravitational acceleration and $h$ is the depth of the water.  For simplicity, we treat $h$ as a constant, which is an appropriate approximation when the wave amplitude is much smaller than $h$.

For the polynomial dispersion, we consider the simplest deviation from the dispersionless case by including a quadratic term in the formula for $c^{2}(k)$.  For the sake of comparison with dispersion relation (\ref{eq:water_dispersion}), the sign of the quadratic term is taken to be negative so that, as there, $c^{2}(k)$ decreases with $k$:
\begin{equation}
c^{2}(k) = c_{0}^{2} \left( 1 - \frac{k^{2}}{3\,k_{0}^{2}} \right) \,.
\label{eq:quartic_dispersion}
\end{equation}
Here, $c_{0}$ is the phase velocity in the long-wavelength limit, and $k_{0}$ indicates the scale at which dispersive effects become important.

\begin{figure}
\subfloat{\includegraphics[width=0.45\columnwidth]{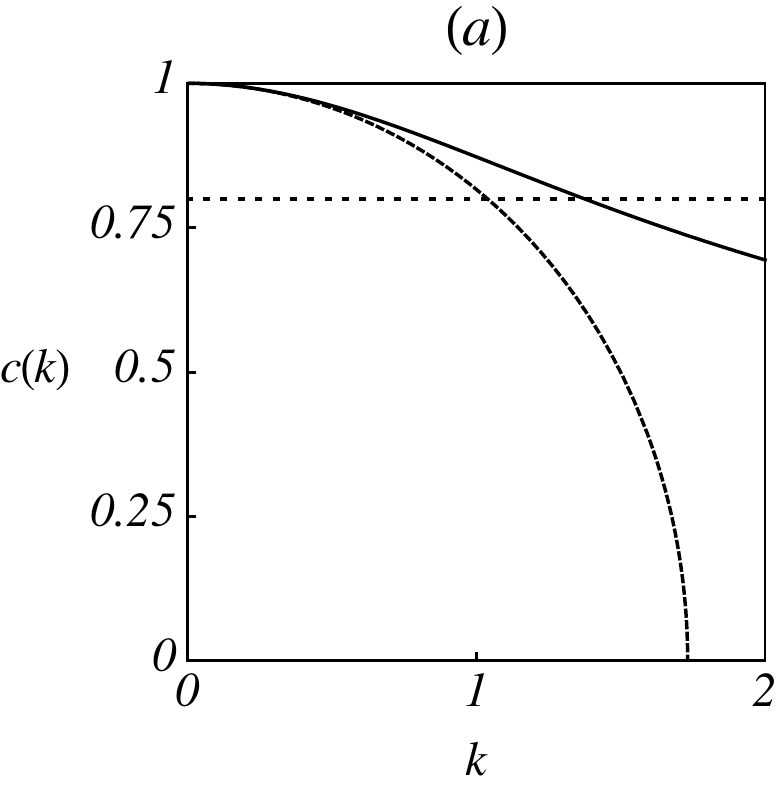}} \, \,\subfloat{\includegraphics[width=0.46\columnwidth]{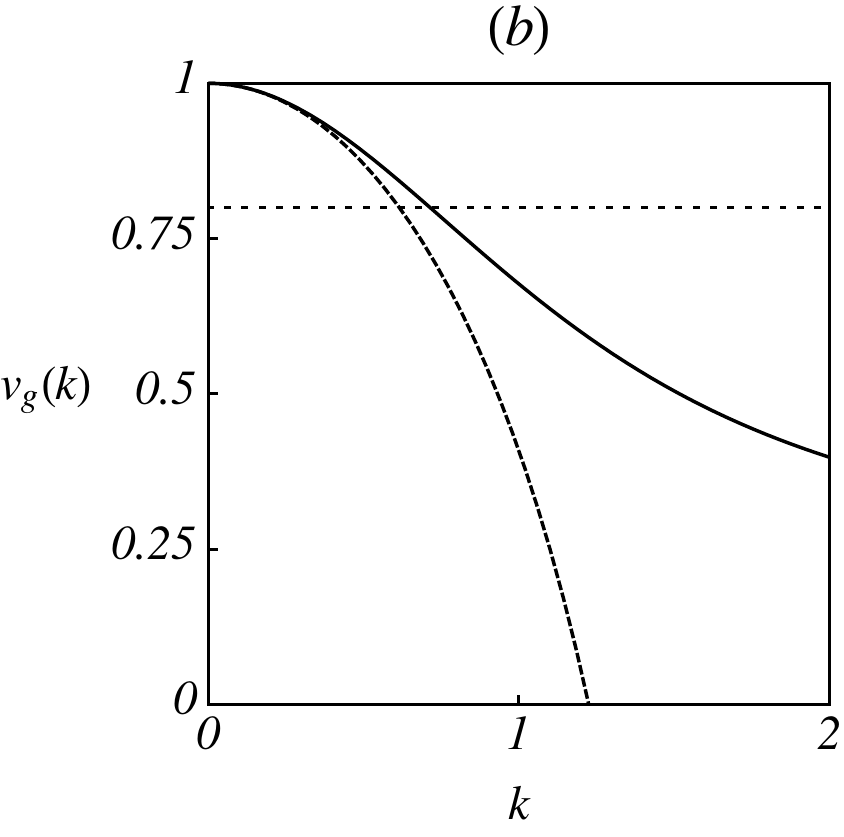}}
\caption{\textsc{Dispersion relations:} \, In $(a)$ are plotted the dimensionless phase velocities $c(k)$, whose squares are given in Eqs. (\ref{eq:normalized_dispersion}); while in $(b)$ are plotted the corresponding group velocities $\partial_{k}\left(c(k)k\right)$.  The solid line plots the first of Eqs. (\ref{eq:normalized_dispersion}), describing surface waves on water; the dashed line shows the second, a polynomial approximation at low $k$.  These are measured with respect to the flow.  The dotted line is at $v=0.8$, which is the magnitude of the asymptotic flow velocity in the velocity profiles of Fig. \ref{fig:velocity}.  These cross at the value of $k$ where, in the stationary frame, the phase or group velocity vanishes, dividing the wavevectors into those which can overcome and those which are dragged with the flow.
\label{fig:dispersion}}
\end{figure}

Dispersion relations (\ref{eq:water_dispersion}) and (\ref{eq:quartic_dispersion}) are chosen to agree in the long-wavelength limit, where standard Hawking radiation is observed.  Since each has two adjustable parameters and their first derivatives vanish automatically at $k=0$, we equate their values and second derivatives at $k=0$.  This is achieved by setting
\begin{alignat}{2}
c_{0}^{2} = gh \,, & \qquad k_{0} = \frac{1}{h} \,.
\label{eq:matching_dispersion_relations}
\end{alignat}
Further simplification occurs by setting these parameters equal to unity, effectively redefining the measure of space and time so that distance is measured in units of $h$ and time in units of $\sqrt{h/g}$.  The resulting dimensionless dispersion relations are simply
\begin{alignat}{3}
c^{2}(k) = \frac{\mathrm{tanh}(k)}{k} \qquad & \mathrm{and} & \qquad c^{2}(k) = 1-\frac{1}{3}k^{2} \,.
\label{eq:normalized_dispersion}
\end{alignat}
The phase velocities $c(k)$ -- along with their corresponding group velocities $\partial_{k}\left(c(k)k\right)$ -- are plotted in Figure \ref{fig:dispersion}.


\subsection{Velocity profile}

\begin{figure}
\includegraphics[width=0.8\columnwidth]{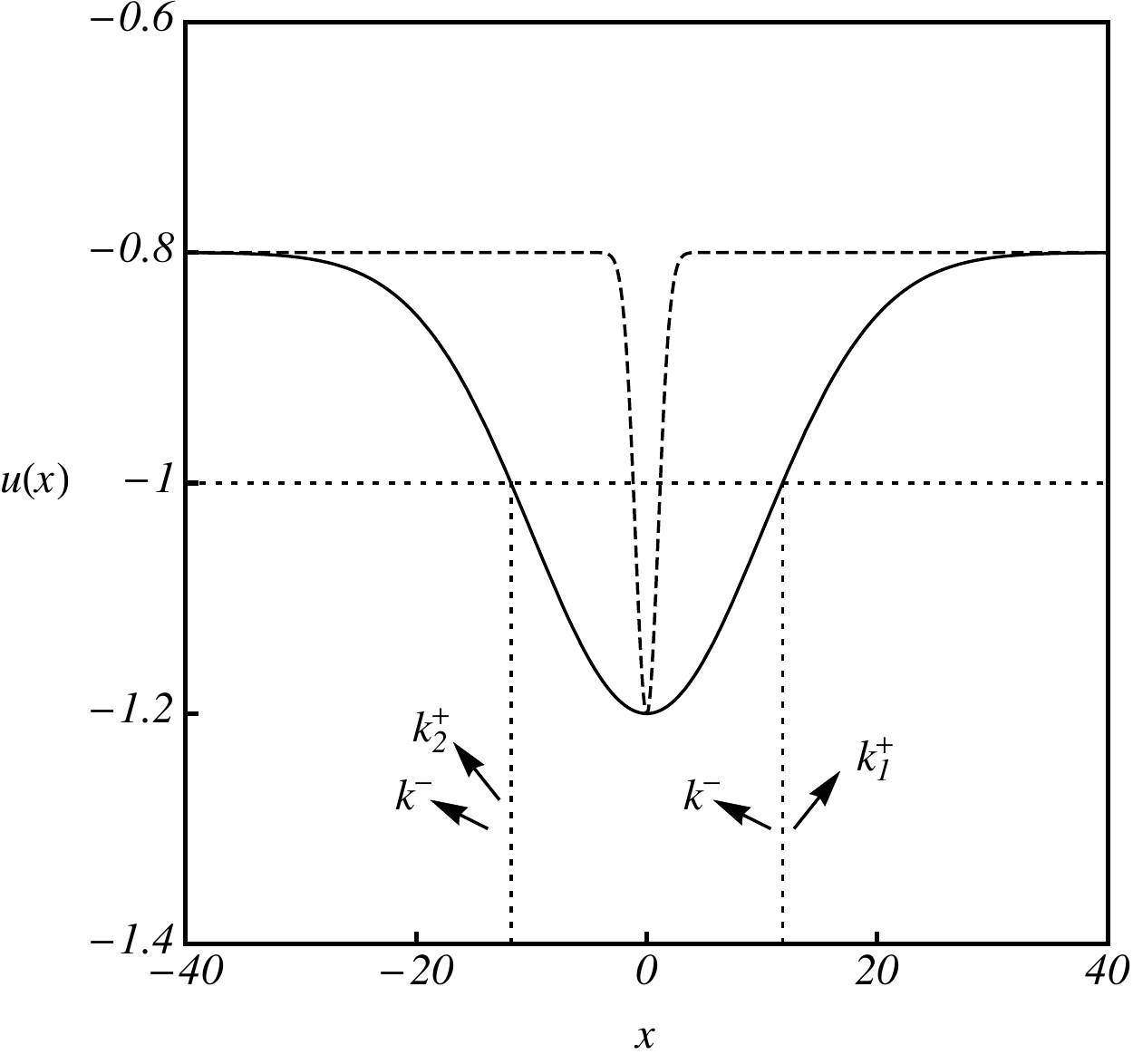}
\caption{\textsc{Velocity profiles:} \, Shown are the flow velocity profiles of Eq. (\ref{eq:velocity_profile}).  The solid line corresponds to the slowly-varying case $a=0.1$, while the dashed line plots the rapidly-varying case where $a=1$.  Dotted lines indicate where the velocity crosses the low-$k$ limiting speed $c_{0}=\sqrt{g h}$, equal to one in dimensionless units.  These points mark the limiting positions of the group velocity horizons, where a wavepacket centred at $k_{1}^{+}$ or $k_{2}^{+}$ is brought to a standstill.  In the region between the horizons, where $u(x)<-1$, only the $k^{-}$ wave can propagate.  The horizon on the right is a black hole horizon, and is associated with production of the $(k_{1}^{+},k^{-})$ pair; while the left horizon is a white hole horizon, associated with production of the $(k_{2}^{+},k^{-})$ pair.  Note that the $k^{-}$ phonons produced at the black hole horizon interact with the white hole horizon, and can stimulate emission of the $(k_{2}^{+},k^{-})$ pair.  This explains the low-frequency peaks of Figs. \ref{fig:temp_hawking}$(b)$ and \ref{fig:temp_non-hawking}$(b)$.
\label{fig:velocity}}
\end{figure}

We shall consider two velocity profiles, differing in their rate of variation.  One will be slowly-varying compared to the scale set by dispersion, and is expected to satisfy Hawking's thermal prediction for the analogue of black hole radiation.  The other will vary rapidly, and we expect to find that Hawking's prediction is no longer valid.  For ease of calculation, we shall consider the modulation of $u(x)$ to have the shape of a Gaussian; its Fourier transform $\tilde{u}(k)$ then has the same shape with inverted standard deviation.  To be precise:
\begin{equation}
u(x) = u_{0} + h \, \mathrm{exp} \left( - \frac{a^{2} x^{2}}{2} \right) \,,
\label{eq:velocity_profile}
\end{equation}
the variation part of which has half-Fourier transforms
\begin{equation}
\mathcal{F}_{L/R}[u-u_{0}](k) = h \, \sqrt{\frac{\pi}{2}} \, a \, \mathrm{exp} \left( - \frac{k^{2}}{2a^{2}} \right) \, \left[ 1 - s_{L/R} \, \mathrm{erf} \left(i \frac{k}{\sqrt{2}\,a}\right) \right] \,,
\label{eq:halfFT_velocity_profile}
\end{equation}
where $s_{L}=-1$ and $s_{R}=1$.  We take $u_{0}=-0.8$ and $h=-0.4$, so that the fluid is left-moving and faster than the maximum wave speed in the inner region (where $c<-1$).  We take $a=0.1$ for the slowly-varying and $a=1$ for the rapidly-varying case.  The velocity profiles are shown in Figure \ref{fig:velocity}.


\subsection{Stationary-frame dispersion and radiation channels}

Frequency is conserved in the stationary frame where the flow velocity is a function only of position.  In this frame, the asymptotic plane waves have frequencies which, due to the flow, are Doppler shifted from their values in the rest frame of the fluid:
\begin{equation}
\Omega^{2} = (\omega - u_{0} k)^{2} = c^{2}(k) k^{2} \,.
\label{eq:Doppler}
\end{equation}
This equation involves $\Omega^{2}$ because there are two branches to the dispersion relation.  These can be grouped into waves which are left-moving and right-moving with respect to the fluid, distinguished by the sign of $\Omega/k$; or they can be grouped into waves of positive and negative norm, which are distinguished by the sign of $\Omega$ \cite{Robertson-2012}.  We will focus on right-moving waves, for which $\Omega/k$ is positive; on the right-moving branch\footnote{The left-moving branch is included in the numerical calculation, but in the system we are considering here, scattering into left-moving waves is small.}, the sign of the norm is the same as the sign of the wavevector $k$.

\begin{figure}
\includegraphics[width=0.8\columnwidth]{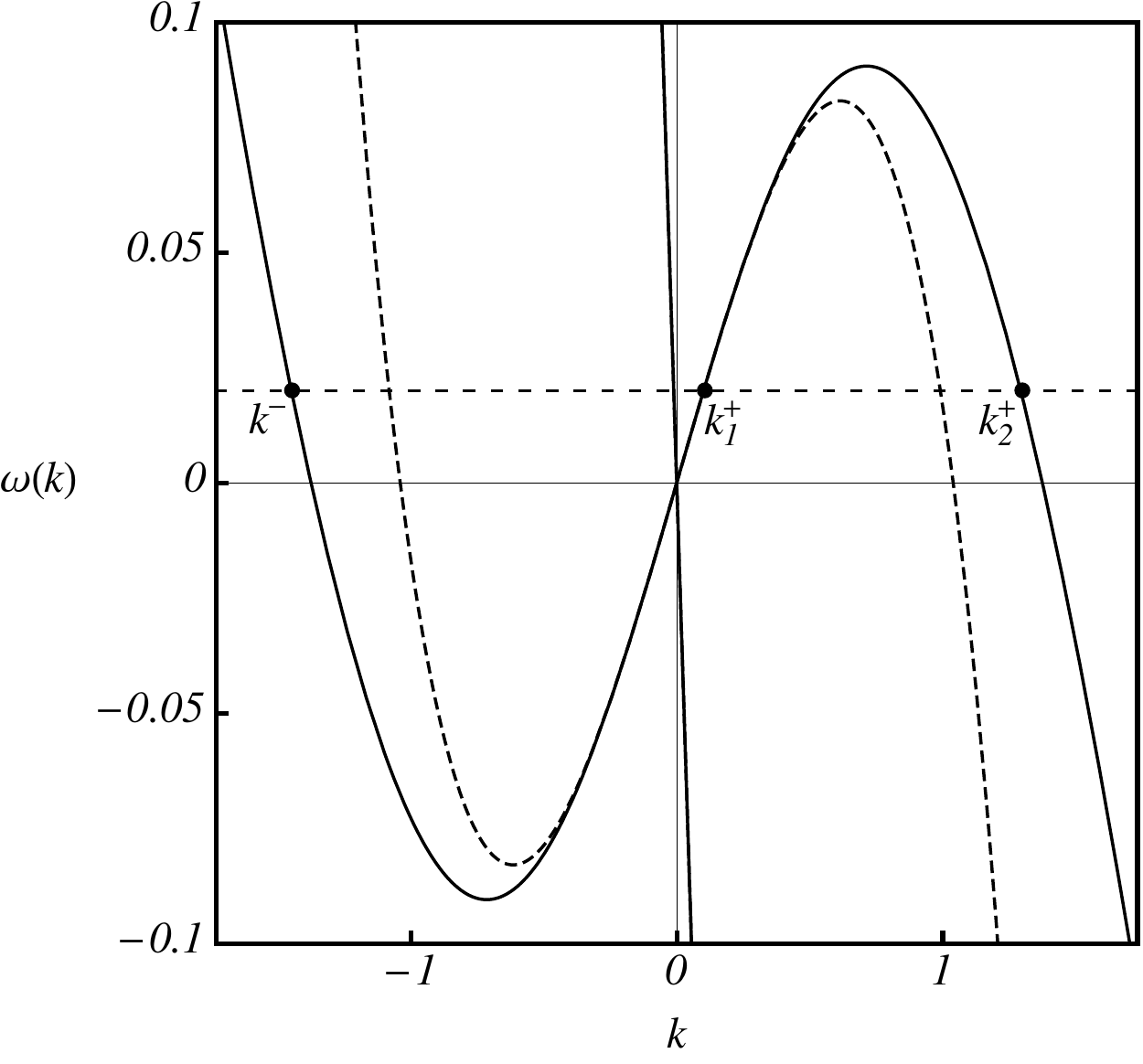}
\caption{\textsc{Dispersion in stationary frame:} \, Here is plotted the frequency as a function of wavevector in the asymptotic regions of the stationary frame, as given by the Doppler formula (\ref{eq:Doppler}).  The solid and dashed curves correspond to the first and second of dispersion relations (\ref{eq:normalized_dispersion}), respectively.  There exist two positive and one negative wavevector at any given frequency below a maximum frequency.  Note that these are all right-moving with respect to the fluid.  (We neglect the left-moving branch, which can be seen running very close to the vertical axis).
\label{fig:dispersion_curve}}
\end{figure}

Figure \ref{fig:dispersion_curve} shows the dispersion relation in the stationary frame.  Drawing a line of fixed $\omega$ (parallel to the $k$-axis), the points at which it crosses the dispersion curve are the allowed asymptotic values of $k$, and therefore represent the asymptotic plane waves that can scatter into each other.  As remarked in Section \ref{sec:Scattering}, Hawking radiation can occur when positive- and negative-norm modes partially scatter into each other; in terms of the dispersion curve, this is when solutions of positive and negative $k$ exist at a fixed value of $\omega$.  What is more, the possible Hawking pairs can consist of any positive $k$ coupled with any negative $k$ of equal frequency.  From Figure \ref{fig:dispersion_curve} we see that (neglecting the left-moving branch) there are two possible Hawking pairs: $\left(k^{+}_{1},k^{-}\right)$ and $\left(k^{+}_{2},k^{-}\right)$.  There are thus two Hawking spectra, each corresponding to one pair.  The $(k_{1}^{+},k^{-})$ pair is the analogue of the standard Hawking radiation.  The $(k_{2}^{+},k^{-})$ pair is associated with the white hole horizon where $u(x)$ climbs back above $-1$; it is different from the standard radiation in that both waves are short wavelength and emitted in the same direction.  (See Fig. \ref{fig:velocity} for an illustration of pair emission at each horizon.)


\subsection{Results}

\begin{figure}
\subfloat{\includegraphics[width=0.465\columnwidth]{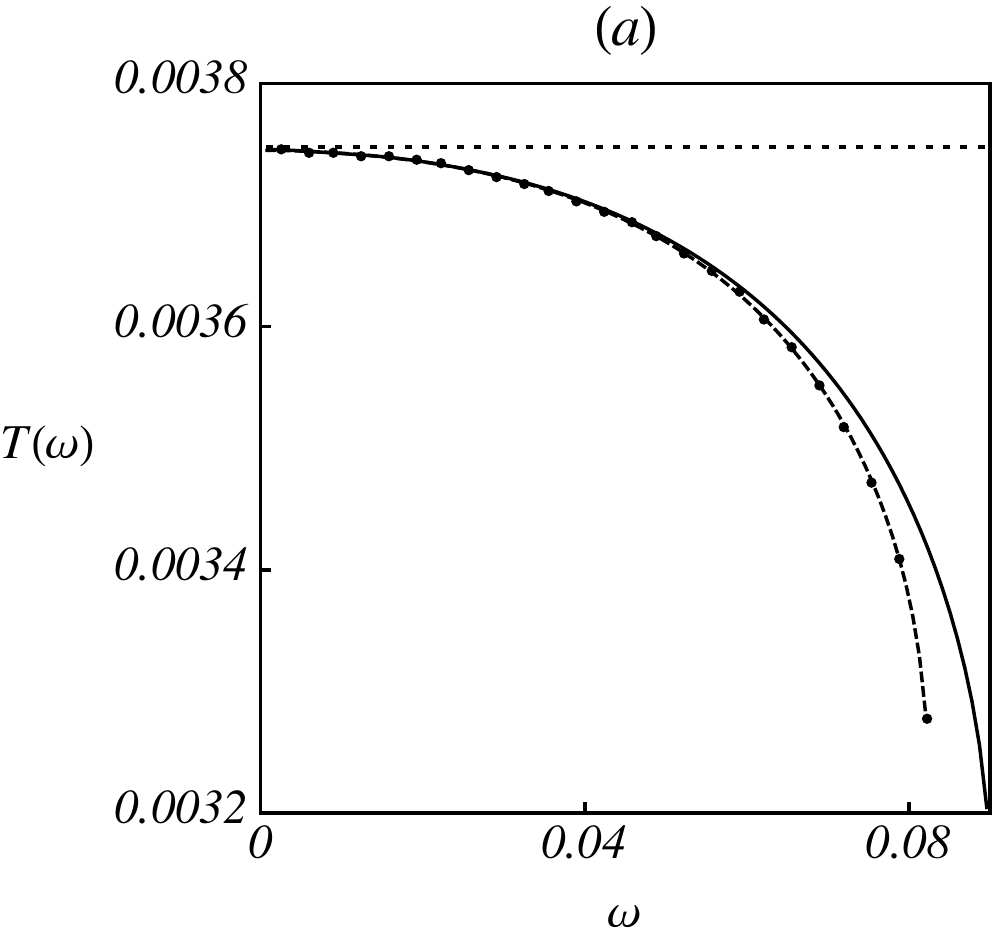}} \, \subfloat{\includegraphics[width=0.45\columnwidth]{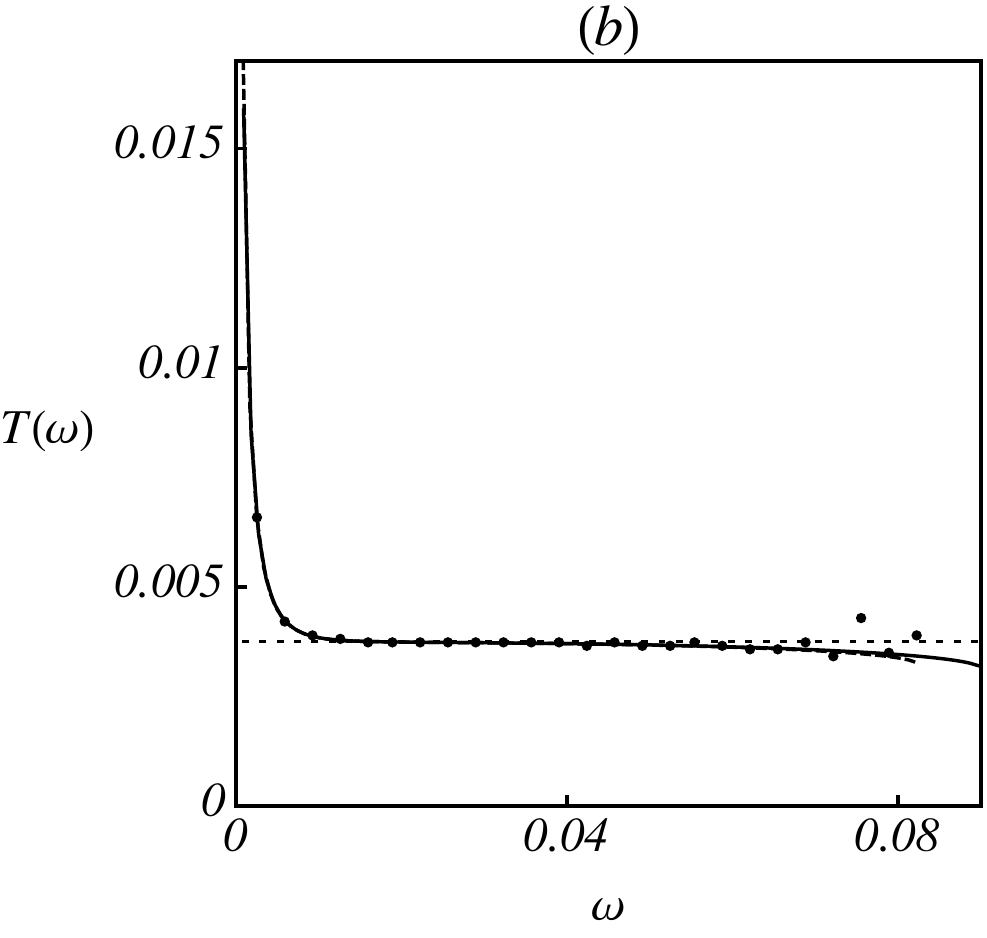}}
\caption{\textsc{Temperature in slowly-varying regime:} \, In $(a)$ is plotted the $\left(k^{+}_{1},k^{-}\right)$ spectrum, and in $(b)$ the $\left(k^{+}_{2},k^{-}\right)$ spectrum, for the velocity profile with $a=0.1$.  Hawking's thermal prediction gives $T_{H}=0.00375$, which is shown by the dotted line.  The solid and dashed lines plot the effective temperature when the dispersion is given by the first and second of Eqs. (\ref{eq:normalized_dispersion}), respectively; the discrete points are the results of solving the position-space ODE for the polynomial dispersion.  The $T$-axis of $(a)$ is taken over a very narrow temperature range, so that Hawking's prediction is a very good approximation for the $\left(k^{+}_{1},k^{-}\right)$ spectrum.  In $(b)$, there is a plateau over which Hawking's prediction is valid, but the effective temperature is much greater than this for low frequencies.
\label{fig:temp_hawking}}
\end{figure}

\begin{figure}
\subfloat{\includegraphics[width=0.45\columnwidth]{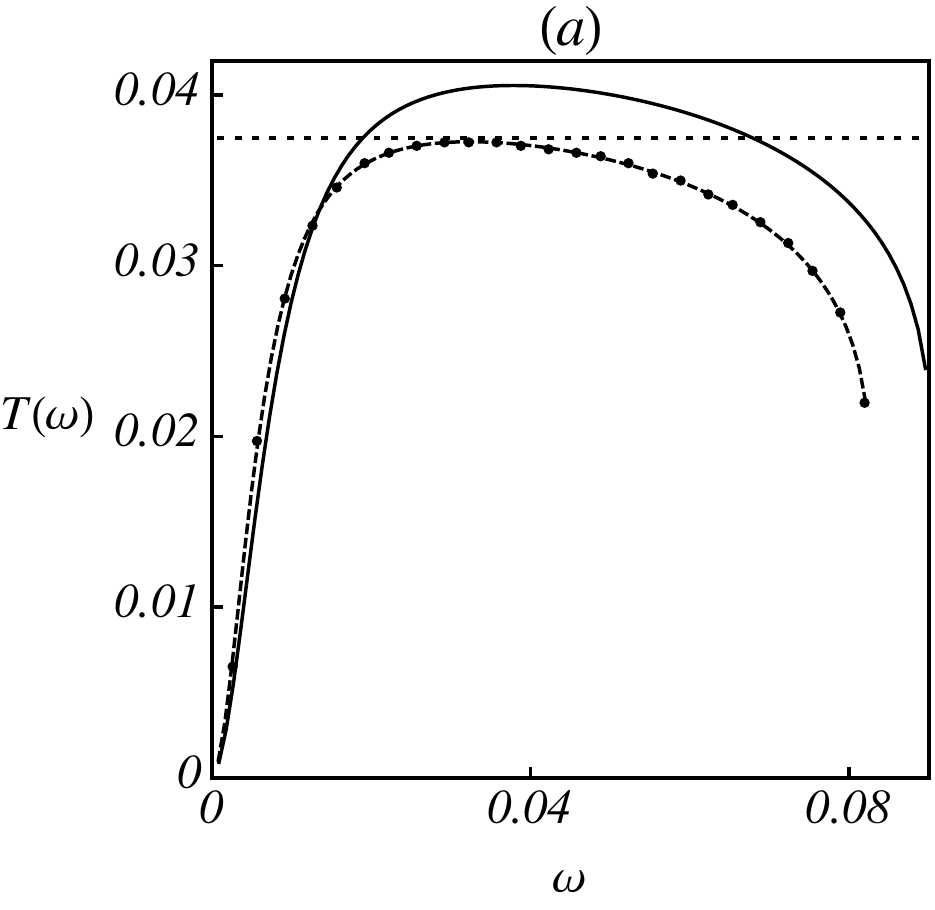}} \, \subfloat{\includegraphics[width=0.45\columnwidth]{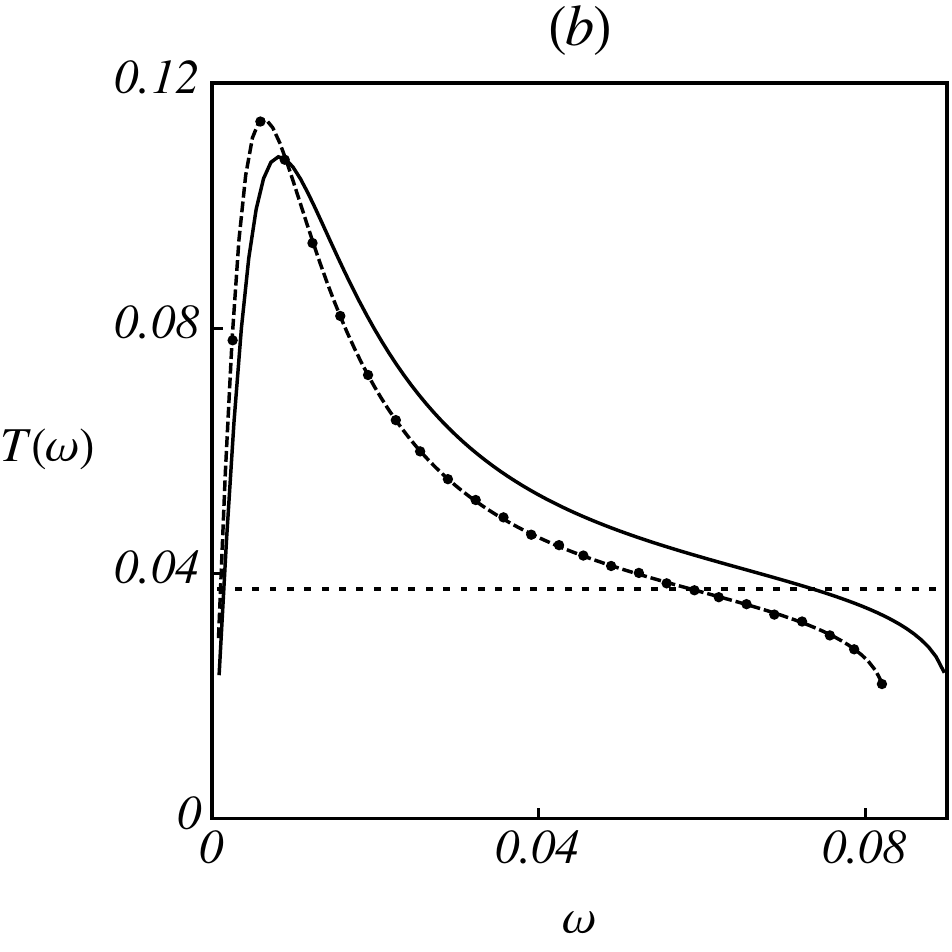}}
\caption{\textsc{Temperature in rapidly-varying regime:} \, In $(a)$ is plotted the $\left(k^{+}_{1},k^{-}\right)$ spectrum, and in $(b)$ the $\left(k^{+}_{2},k^{-}\right)$ spectrum, for the velocity profile with $a=1$.  Hawking's thermal prediction gives $T_{H}=0.0375$, which is shown by the dotted line.  The solid and dashed lines plot the effective temperature when the dispersion is given by the first and second of Eqs. (\ref{eq:normalized_dispersion}), respectively; the discrete points are the results of solving the position-space ODE for the polynomial dispersion.  Note that Hawking's prediction is much less valid here than in the slowly-varying case.
\label{fig:temp_non-hawking}}
\end{figure}

Being scalar and massless, a truly thermal spectrum of the field would have (in dimensionless units) the occupation number
\begin{equation}
\left| \beta_{\omega} \right|^{2} = \frac{1}{e^{\omega/T}-1} \,.
\label{eq:thermal_spectrum}
\end{equation}
Since the spectrum itself varies over many orders of magnitude, it is convenient to plot the frequency-dependent effective temperature
\begin{equation}
T(\omega) = \frac{\omega}{\mathrm{ln} \left( 1 + 1/|\beta_{\omega}|^{2} \right) } \,.
\label{eq:effective_temperature}
\end{equation}
Figures \ref{fig:temp_hawking} and \ref{fig:temp_non-hawking} plot the effective temperature of the Hawking spectra in the slowly-varying ($a=0.1$) and rapidly-varying ($a=1$) cases, respectively.  On each plot are given the spectra for both the realistic water wave dispersion and the artificial polynomial dispersion (see Eqs. (\ref{eq:normalized_dispersion})).  Also shown as discrete points are the results from solving the position-space ODE (see Eq. (\ref{eq:dispersive_wave_eqn})) for the polynomial dispersion.  The fact that these agree with the curves calculated using the integral equation method discussed in this paper show that it is valid, and it is robust with respect to variation of the velocity profile.  Moreover, because of the relative wideness of the velocity profile in the slowly-varying case, the integration range of the ODE must be taken to be large, which both increases the computing time and decreases the accuracy, as can be seen from the discrepancies of some of the high-frequency points in Figure \ref{fig:temp_hawking}$(b)$.

Hawking's original thermal prediction \cite{Hawking-1974,Hawking-1975}, when applied to a dispersionless fluid with arbitrary velocity profile \cite{Robertson-2012}, is that the spectrum is exactly thermal with temperature proportional to the derivative of the velocity profile at the horizon:
\begin{equation}
T_{H} = \frac{u^{\prime}\left(x_{h}\right)}{2\pi} \,.
\label{eq:Hawking_prediction}
\end{equation}
The predictions of Eq. (\ref{eq:Hawking_prediction}) are also shown in Figs. \ref{fig:temp_hawking} and \ref{fig:temp_non-hawking}.  We note that, in the slowly-varying regime, the spectra are (apart from a low-frequency peak in the $\left(k^{+}_{2},k^{-}\right)$ spectrum) well-approximated by Eq. (\ref{eq:Hawking_prediction}), and much less so in the rapidly-varying case.

\begin{figure}
\subfloat{\includegraphics[width=0.485\columnwidth]{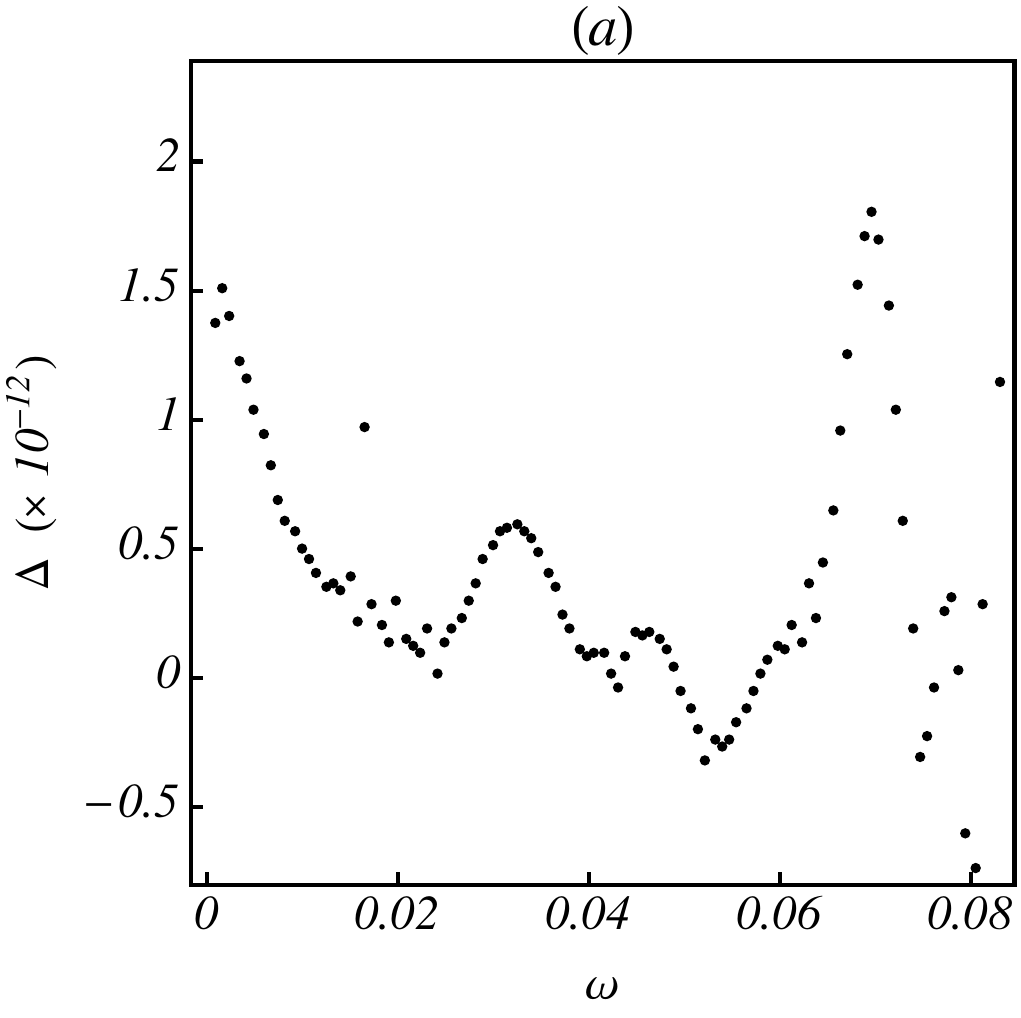}} \, \subfloat{\includegraphics[width=0.45\columnwidth]{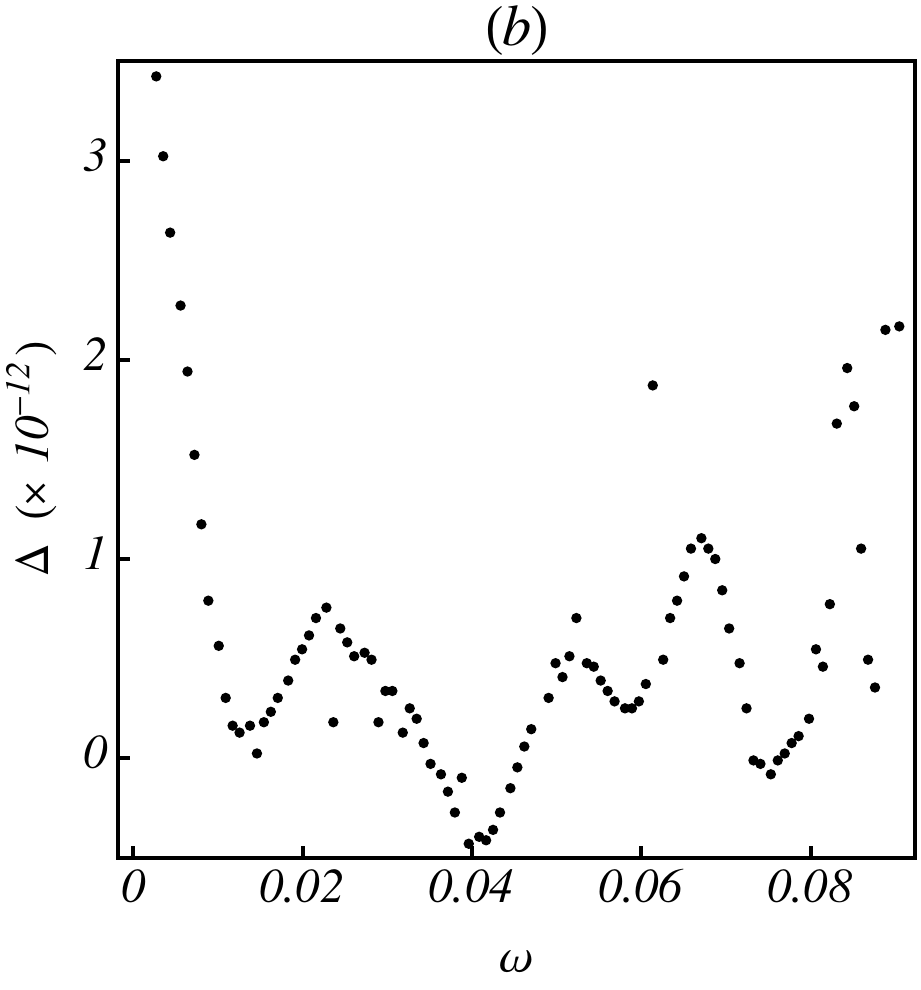}}
\caption{\textsc{Accuracy of norm conservation}: \, Here is plotted $\Delta$, the difference in norm between ingoing and outgoing waves for the $k_{1}^{+}$ out-mode; equivalently, it is Eq. (\ref{eq:norm_conservation}) minus $1$.  For the exact solution, it should be exactly zero.  In $(a)$, it is plotted for the polynomial dispersion, the second of Eqs. (\ref{eq:normalized_dispersion}); and in $(b)$, for the first of Eqs. (\ref{eq:normalized_dispersion}) describing water wave dispersion.  In both cases, the slowly-varying case ($a=0.1$) was considered, and in the discretisation of the integration grid, we have used $k_{\mathrm{max}}=4$ and $M=100$.  It can be seen in both cases that $\Delta$ is very small, on the order of $10^{-12}$, and is of the same order of magnitude for both dispersion relations.  For the rapidly-varying case ($a=1$), $\Delta$ is between one and two orders of magnitude smaller than it is here; and in all cases, $\Delta$ calculated for the $k_{2}^{+}$ out-mode is also of the same order of magnitude.  This shows that (at least for the dispersion and velocity profiles used here), the integral method produces very accurate results.
\label{fig:Delta}}
\end{figure}

In Figure \ref{fig:Delta} we plot the difference in norm between the ingoing and outgoing waves for the $k_{1}^{+}$ out-mode; that is, with the decomposition of Eq. (\ref{eq:out-in_decomp}), we plot
\begin{equation}
\Delta_{\omega} = \sum_{j} \left| \alpha_{\omega} \right|^{2} - \sum_{j} \left| \beta_{\omega} \right|^{2} - 1 \,.
\label{eq:Delta_defn}
\end{equation}
This should be identically zero for the exact solution.  Its value thus gives an indication of the accuracy of the numerical calculation of the scattering amplitudes.  In all the results of Figs. \ref{fig:temp_hawking}-\ref{fig:Delta}, we used a discretised integration grid with $M=100$ points between $\pm k_{\mathrm{max}}$, where $k_{\mathrm{max}}=4$ for the slowly-varying case and $k_{\mathrm{max}}=20$ for the rapidly-varying case.  In all cases, $\Delta_{\omega}$ is found to be very small, $\lesssim 10^{-12}$, which is much smaller than any of the calculated scattering amplitudes except {\it very} near the upper edge of the Hawking spectrum where it falls rapidly to zero.  This shows that, at least for the dispersion and velocity profiles considered here, the integral method produces very accurate results.


\section{Conclusion
\label{sec:Conclusion}}

In this paper we have presented a new method for the calculation of the scattering matrix of dispersive waves in a one-dimensional background, with particular emphasis on its application to analogue Hawking radiation.  The method solves an integral equation in Fourier space, utilizing the analytical properties of the half-Fourier transforms of the solution, and is amenable to standard and efficient numerical techniques.  Unlike direct solution of an ODE in position space, which restricts the dispersion relation to a polynomial of relatively low degree, the method allows for arbitrary dispersion.  We have shown that it agrees with the solutions of the ODE when the latter can be solved, and we have shown that it solves for a more complicated dispersion relation with no additional difficulty.

The main restriction we have imposed in this paper is that the asymptotic values of the background are equal in the left- and right-hand regions.  This assumption greatly simplifies the analytical development of the method.  While this can describe many black hole analogue systems, such as Hawking radiation from pulses in optical fibres \cite{Philbin-et-al-2008}, it is not the most general case.  The generalisation to different asymptotic values of the background are presented in Part II.


\begin{acknowledgments}
SR is grateful to South China Normal University and the Weizmann Institute of Science for hospitality and financial support.
\end{acknowledgments}


\newpage
\appendix
\appendixpage
\numberwithin{equation}{section}

\section{Splitting the kernel into half-plane analytic parts
\label{app:splitting_kernel}}

In this appendix we shall briefly discuss how the integral kernel $K(k,k^{\prime})$ is split into two parts which are analytic on the upper and lower half-planes of complex $k^{\prime}$, as performed in Eq. (\ref{eq:splitting_kernel}).

The kernel derives from the position-dependent terms of the wave equation; in the case of Eq. (\ref{eq:dispersive_wave_eqn}), it is determined by the flow velocity profile $u(x)$.  When the Fourier transform of the wave equation is taken, these terms become convolutions.  The first line of Eq. (\ref{eq:fluid_kernel}) is the expression for $K(k,k^{\prime})$ obtained directly from these convolutions, with the dummy variable $k^{\prime}$ (integrated over in the resulting integral equation) appearing only in the arguments of Fourier transforms of $u$ and various derivatives of $u$.  Note that this expression remains valid when the $\delta$ functions in $\mathcal{F}[u]$ and $\mathcal{F}[u^{2}]$ are extracted, with $u$ and $u^{2}$ reinterpreted as $u-u_{0}$ and $u^{2}-u_{0}^{2}$, respectively.  Since $u(x)$ is asymptotically bounded, the half-plane analytic parts of $K(k,k^{\prime})$ are straightforwardly found by splitting each of the Fourier transforms into half-Fourier transforms.  So, Eq. (\ref{eq:splitting_kernel}) holds with
\begin{multline}
K_{L/R}(k,k^{\prime}) = \frac{1}{2\pi} \left[ 2\omega k \mathcal{F}_{L/R}[u](k-k^{\prime}) + i\omega \mathcal{F}_{L/R}[\partial_{x}u](k-k^{\prime}) \right. \\
\left. -k^{2} \mathcal{F}_{L/R}[u^{2}](k-k^{\prime}) - i k \mathcal{F}_{L/R}[\partial_{x}u^{2}](k-k^{\prime}) \right] \, ,
\label{eq:kernel_halfFTs}
\end{multline}
where $K_{L}(k,k^{\prime})$ is analytic and vanishes like $1/k^{\prime}$ on the lower half $k^{\prime}$-plane, and $K_{R}(k,k^{\prime})$ behaves similarly on the upper-half $k^{\prime}$-plane.

One notable exception occurs when $u$ is discontinuous at $x=0$, for then the derivatives $\partial_{x}u$ and $\partial_{x}u^{2}$ contain $\delta$ functions at $x=0$.  This $\delta$ function contributes a purely constant term to their Fourier transforms.  In this case, while $K_{L/R}(k,k^{\prime})$ can still be defined with the same analyticity properties by using the limits $0\pm \epsilon$ in the definitions of the half-Fourier transforms -- thus avoiding integration over the $\delta$ functions -- Eq. (\ref{eq:splitting_kernel}) is no longer true, being replaced by
\begin{equation}
K(k,k^{\prime}) = K_{L}(k,k^{\prime}) + K_{R}(k,k^{\prime}) + \frac{i}{2\pi} \left(u^{+}-u^{-}\right) \left( \omega - \left(u^{+}+u^{-}\right) k \right) \,,
\end{equation}
where $u^{\pm} \equiv u\left(0 \pm \epsilon\right)$.  Equation (\ref{eq:left-right_integral_equation}) would also contain an additional term proportional to $\phi(x=0)$.  While we have focused on the example kernel (\ref{eq:fluid_kernel}), similar considerations will hold for kernels of other wave equations.  In this paper, we assume that the background is smooth enough at $x=0$ such that Eq. (\ref{eq:splitting_kernel}) is true.  We also note in passing that a discontinuity at any other point does not lead to this complication, since the Fourier transform of a $\delta$ function located at an arbitrary position $x_{0}$ is $e^{-ikx_{0}}$, which for $x_{0} \neq 0$ is analytic {\it and} exponentially vanishing on one or other half-plane according to the sign of $x_{0}$ and can be included in one of the half-kernels $K_{L/R}(k,k^{\prime})$.

The second line of Eq. (\ref{eq:fluid_kernel}) derives from the relation
\begin{equation}
\mathcal{F}[\partial_{x}f](k) = ik \mathcal{F}[f](k)
\label{eq:derivative_FT}
\end{equation}
between the Fourier transform of a function $f$ that vanishes asymptotically and the Fourier transform of its derivative.  It simplifies the first line, replacing four Fourier transforms with two.  However, we must be cautious when attempting to simplify the decomposition of Eq. (\ref{eq:kernel_halfFTs}) in a similar fashion.  Equation (\ref{eq:derivative_FT}) holds between {\it full} Fourier transforms, because the boundary terms left over from integration by parts are then at $\pm \infty$, where the integrand vanishes.  By contrast, with a {\it half}-Fourier transform, one of the boundary terms is evaluated at $x=0$, and must be included in the corresponding relation.  Another way to see the necessary modification to Eq. (\ref{eq:derivative_FT}) for half-Fourier transforms is to note that it holds exactly for $\theta\left(\mp x\right) f(x)$.  Then the right-hand side becomes exactly $ik \mathcal{F}_{L/R}[f](k)$, but a $\delta$ function term, $\mp f\left(0\right) \delta(x)$, must be added to $\partial_{x}f$, which becomes simply $\mp f\left(0\right)$ upon Fourier transformation.  (We are assuming $f$ is continuous at $x=0$ so that $f(0)$ is unambiguously defined.)  Thus, the half-Fourier transform equivalent of Eq. (\ref{eq:derivative_FT}) is
\begin{equation}
\mathcal{F}_{L/R}[\partial_{x}f](k) = ik \mathcal{F}_{L/R}[f](k) \pm f\left(0\right) \,.
\label{eq:derivative_halfFT}
\end{equation}
and upon applying Eq. (\ref{eq:derivative_halfFT}) to Eq. (\ref{eq:kernel_halfFTs}), we find
\begin{multline}
K_{L/R}(k,k^{\prime}) = \frac{1}{2\pi} \left[ \omega (k+k^{\prime}) \mathcal{F}_{L/R}[u](k-k^{\prime}) - k k^{\prime} \mathcal{F}_{L/R}[u^{2}](k-k^{\prime}) \right. \\
\left. \pm i u(0) \left( \omega - u(0) k\right) \right] \,.
\label{eq:kernel_halfFTs_noD}
\end{multline}
Again, though this is for the specific example of Eq. (\ref{eq:fluid_kernel}), a similar modification will hold for other kernels.


\section{Matrix multiplication suitable for numerics}

By construction, the solution to the integral equation depends critically on the solutions to the dispersion relation ({\it i.e.} the zeros of $g(k)$) at the given value of $\omega$.  If, as is usually the case, we wish to calculate a radiation spectrum over a large frequency range, it becomes expedient to find the most economical way of constructing the scattering matrix.  While matrix operations such as multiplication and inversion can be done efficiently by standard algorithms, entering initial values into large matrices can be time-consuming.  It is best to do this as little as possible, and to form matrices derived from initial ones via standard matrix operations.

The kernel $K(k,k^{\prime})$ can contain powers of $\omega$, which come from the time derivatives in the wave equation when the harmonic time dependence $e^{-i\omega t}\phi_{\omega}(x)$ is assumed.  A typical wave equation is of finite order in $\partial_{t}$, so that the dependence of the kernel on $\omega$ is simply a polynomial, the coefficients of which are $\omega$-independent: they can be initialized {\it once} for a particular spectrum.  Upon discretization, the coefficients become matrices $\hat{K}^{\delta\delta,(p)}_{nm} = K^{(p)}\left(k_{n},k_{m}\right)$, where we have introduced a hat to indicate a matrix to be stored or calculated numerically, and the label $\delta$ is used to indicate that the arguments are elements of the discretised integration grid.  Then, for each $\omega$,
\begin{equation}
\hat{K}^{\delta\delta}_{nm} = \sum_{p} \omega^{p} \, \hat{K}^{\delta\delta,(p)}_{nm} \,.
\end{equation}

Recalling that we use subscripts $m$ and $n$ to denote $k$-values on the discretised integration grid -- on which there are $M$ points -- and subscripts $i$ and $j$ to denote the real roots of $g(k)$ -- of which there are $N \ll M$ -- then the matrix $\hat{\bar{K}}^{\delta\delta}_{nm}$ is given explicitly by
\begin{equation}
\hat{\bar{K}}^{\delta\delta}_{nm} = \bar{K}\left(k_{n},k_{m}\right) = \frac{K\left(k_{n},k_{m}\right)}{g\left(k_{n}\right)} - \sum_{j=1}^{N} \frac{K\left(k_{j},k_{m}\right)}{\left( k_{n}-k_{j} \right) g^{\prime}\left(k_{j}\right)} \,.
\label{eq:kbar_matrix}
\end{equation}
We define the following matrices:
\begin{alignat}{3}
\hat{g}^{-1}_{nm} = \frac{1}{g\left(k_{n}\right)} \, \delta_{nm} \,, & \qquad \hat{\Delta}_{nj} = \frac{1}{\left(k_{n}-k_{j}\right) \, g^{\prime}\left(k_{j}\right)} \,, & \qquad \hat{K}^{r \delta}_{jm} = K\left(k_{j},k_{m}\right) \, ,
\label{eq:numerics_g_Delta_K}
\end{alignat}
of dimensions $M \times M$, $M \times N$ and $N \times M$, respectively; the label $r$ has also been introduced to indicate that the corresponding argument is a root of the dispersion relation, and not an element of the integration grid.  Now, $\hat{g}^{-1}$ being diagonal, the number of non-zero elements of matrices (\ref{eq:numerics_g_Delta_K}) is of order $M$ rather than $M^{2}$, and they can be efficiently calculated for each point of the spectrum.  In terms of these, the $M \times M$ matrix $\hat{\bar{K}}^{\delta\delta}$ can be expressed as
\begin{equation}
\hat{\bar{K}}^{\delta\delta} = \hat{g}^{-1} \cdot \hat{K}^{\delta\delta} - \hat{\Delta} \cdot \hat{K}^{r\delta} \,,
\label{eq:Kbar_matrix}
\end{equation}
and the $M \times M$ matrix $\hat{V}$ is then simply
\begin{equation}
\hat{V} = \left[ \mathbb{1}_{M} + \Delta k \cdot \hat{\bar{K}}^{\delta\delta} \right]^{-1} \,.
\end{equation}
The remaining matrices can also be efficiently calculated for each value of $\omega$:
\begin{alignat}{3}
\hat{g}^{\prime}_{ij} = g^{\prime}\left(k_{i}\right) \, \delta_{ij} \,, & \qquad \left[\hat{K}^{rr}_{L/R}\right]_{ij} = K_{L/R}\left(k_{i},k_{j}\right) \,, & \qquad \left[\hat{K}^{\delta r}_{L/R}\right]_{nj} = K_{L/R}\left(k_{n},k_{j}\right) \,,
\label{eq:numerics_gr_Krr_Kdr}
\end{alignat}
of dimensions $N \times N$, $N \times N$ and $M \times N$, respectively; in terms of these we have
\begin{equation}
\hat{\bar{K}}_{L/R}^{\delta r} = \hat{g}^{-1} \cdot \hat{K}_{L/R}^{\delta r} - \hat{\Delta} \cdot \hat{K}_{L/R}^{rr} \,,
\end{equation}
which is analogous to Eq. (\ref{eq:Kbar_matrix}) with the second argument replaced by a root of the dispersion relation and $K$ replaced by its left or right component.  Finally, we can write the $N \times N$ matrices $\hat{\mathcal{M}}_{L}$ and $\hat{\mathcal{M}}_{R}$ of Eq. (\ref{eq:left-right_matrices_discretized}) entirely in terms of the matrices defined above:
\begin{subequations}\begin{eqnarray}
\hat{\mathcal{M}}_{L} & = & -\frac{1}{2\pi\,i} \, \hat{g}^{\prime} + \hat{K}^{rr}_{L} - \Delta k \cdot \hat{K}^{r\delta} \cdot \hat{V} \cdot \hat{\bar{K}}^{\delta r}_{L} \,, \\
\hat{\mathcal{M}}_{R} & = & \frac{1}{2\pi\,i} \, \hat{g}^{\prime} + \hat{K}^{rr}_{R} - \Delta k \cdot \hat{K}^{r\delta} \cdot \hat{V} \cdot \hat{\bar{K}}^{\delta r}_{R} \,.
\end{eqnarray}\end{subequations}



\newpage

\end{document}